# Connecting the dots between mechanosensitive channel abundance, osmotic shock, and survival at single-cell resolution

Griffin Chure[a,†], Heun Jin Lee[b,†], and Rob Phillips[a, b,]

[a] Department of Biology and Biological Engineering, California Institute of Technology, Pasadena, CA, USA

[b] Department of Physics, California Institute of Technology, Pasadena, California, USA

† contributed equally

**Running title:** MscL copy number and cell survival after osmotic shock

\* Send correspondence to `phillips@pboc.caltech.edu`

## Abstract

Rapid changes in extracellular osmolarity are one of many insults microbial cells face on a daily basis. To protect against such shocks, *Escherichia coli* and other microbes express several types of transmembrane channels which open and close in response to changes in membrane tension. In *E. coli*, one of the most abundant channels is the mechanosensitive channel of large conductance (MscL). While this channel has been heavily characterized through structural methods, electrophysiology, and theoretical modeling, our understanding of its physiological role in preventing cell death by alleviating high membrane tension remains tenuous. In this work, we examine the contribution of MscL alone to cell survival after osmotic shock at single cell resolution using quantitative fluorescence microscopy. We conduct these experiments in an *E. coli* strain which is lacking all mechanosensitive channel genes save for MscL whose expression is tuned across three orders of magnitude through modifications of the Shine-Dalgarno sequence. While theoretical models suggest that only a few MscL channels would be needed to alleviate even large changes in osmotic pressure, we find that between 500 and 700 channels per cell are needed to convey upwards of 80% survival. This number agrees with the average MscL copy number measured in wild-type *E. coli* cells through proteomic studies and quantitative Western blotting. Furthermore, we observe zero survival events in cells with less than 100 channels per cell. This work opens new questions concerning the contribution of other mechanosensitive channels to survival as well as regulation of their activity.

## Importance

Mechanosensitive (MS) channels are transmembrane protein complexes which open and close in response to changes in membrane tension as a result of osmotic shock. Despite extensive biophysical characterization, the contribution of these channels to cell survival remains largely unknown. In this work, we use quantitative video microscopy to measure the abundance of a single species of MS channel in single cells followed by their survival after a large osmotic shock. We observe total death of the population with less than 100 channels per cell and determine that approximately 500 - 700 channels are needed for 80% survival. The number of channels we find to confer nearly full survival is consistent with the counts of the number of channels in wild type cells in several earlier studies. These results prompt further studies to dissect the contribution of other channel species to survival.

## Introduction

Changes in the extracellular osmolarity can be a fatal event for the bacterial cell. Upon a hypo-osmotic shock, water rushes into the cell across the membrane, leaving the cell with no choice but to equalize the pressure. This equalization occurs either through damage to the cell membrane (resulting in death) or through the regulated flux of water molecules through transmembrane protein channels (Fig 1A). Such proteinaceous pressure release valves have been found across all domains of life, with the first bacterial channel being described in 1987 (1). Over the past thirty years, several more channels have been discovered, described, and (in many cases) biophysically characterized. *E. coli*, for example, has seven of these channels (one MscL and six MscS homologs) which have varied conductance, gating mechanisms, and expression levels. While they have been the subject of much experimental and theoretical dissection, much remains a mystery with regard to the roles their abundance and interaction with other cellular processes play in the greater context of physiology (2–8).

Of the seven channels in *E. coli*, the mechanosensitive channel of large conductance (MscL) is one of the most abundant and the best characterized. This channel has a large conductance (3 nS) and mediates the flux of water molecules across the membrane via a ~3 nm wide pore in the open state (9, 10). Molecular dynamics simulations indicate that a single open MscL channel permits the flux of $4 \times 10^9$ water molecules per second, which is an order of magnitude larger than a single aquaporin channel (BNID 100479) (11, 12). This suggests that having only a few channels per cell could be sufficient to relieve even large changes in membrane tension. Electrophysiological experiments have suggested a small number of channels per cell (13, 14), however, more recent approaches using quantitative western blotting, fluorescence microscopy, and proteomics have measured several hundred MscL per cell (3, 15, 16). To further complicate matters, the expression profile of MscL appears to depend on growth phase, available carbon source, and other environmental challenges (3, 16, 17). While there are likely more than just a few channels per cell, why cells seem to need so many and the biological rationale behind their condition-dependent expression both remain a mystery.

While their biochemical and biophysical characteristics have received much attention, their connection to cell survival is understudied. Drawing such a direct connection between channel copy number and survival requires quantitative *in vivo* experiments. To our knowledge, the work presented in van den Berg et al. 2016 (8) is the first attempt to simultaneously measure channel abundance and survivability for a single species of mechanosensitive channel. While the measurement of channel copy number was performed at the level of single cells using super-resolution microscopy, survivability after a hypo-osmotic shock was assessed in bulk plating assays which rely on serial dilutions of a shocked culture followed by counting the number of resulting colonies after incubation. Such bulk assays have long been the standard for querying cell viability after an osmotic challenge. While they have been highly informative, they reflect only the mean survival rate of the population, obfuscating the variability in survival of the population. The stochastic nature of gene expression results in a noisy distribution of MscL channels rather than a single value, meaning those found in the long tails of the distribution have quite different survival rates than the mean but are lost in the final calculation of survival probability.

In this work, we present an experimental system to quantitatively probe the interplay between MscL copy number and survival at single-cell resolution, as is seen in Fig. 1B. We generated an *E. coli* strain in which all seven mechanosensitive channels had been deleted from the chromosome followed by a chromosomal integration of a single gene encoding an MscL-super-folder GFP (sfGFP) fusion protein. To explore copy number regimes beyond those of the wild-type expression level, we modified the Shine-Dalgarno sequence of this integrated construct allowing us to cover nearly three decades of MscL copy number. To probe survivability, we exposed cells to a large hypo-osmotic shock at controlled rates in a flow cell under a microscope, allowing the observation of the single-cell channel copy number and the resulting survivability of single cells. With this large set of single cell measurements, we approach the calculation of survival probability in a manner that is free of binning bias which allows the reasonable extrapolation of survival probability to copy numbers outside of the observed range. In addition, we show that several hundred channels are needed to convey high rates of survival and observe a minimum number of channels needed to permit any degree of survival.

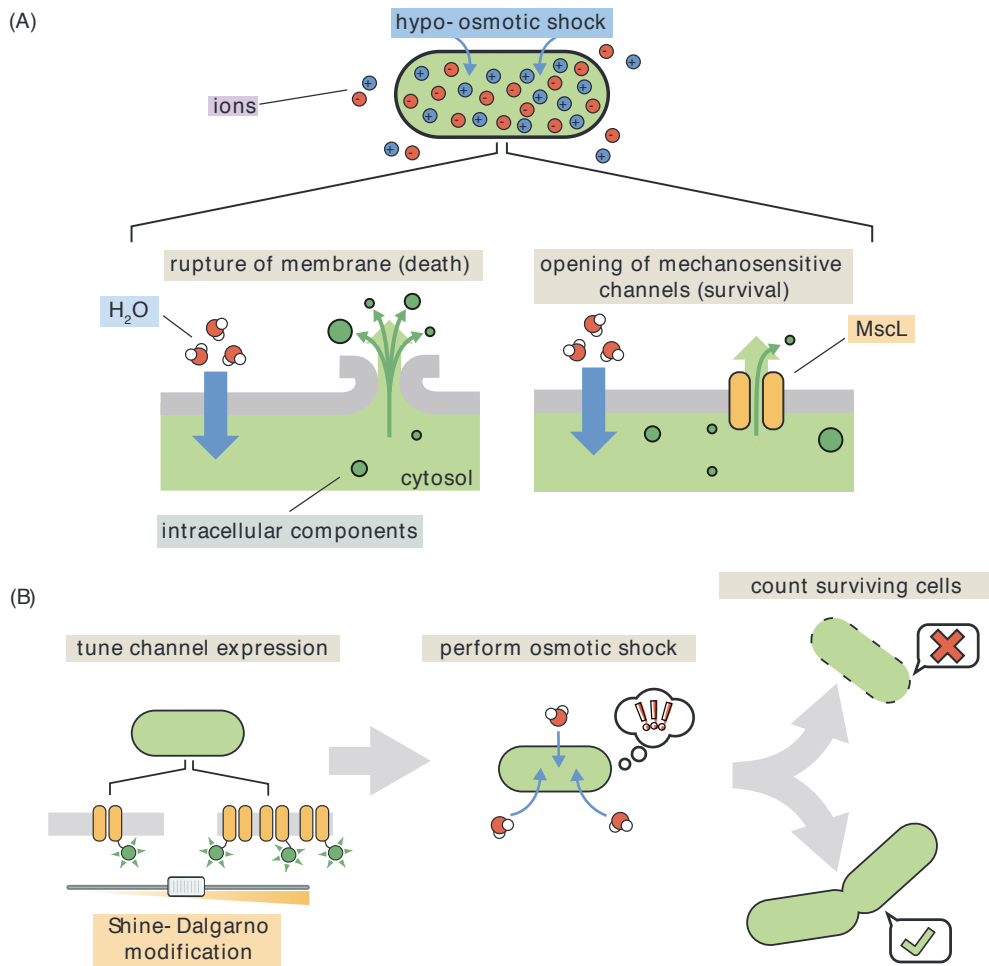

FIG 1 Role of mechanosensitive channels during hypo-osmotic shock. (A) A hypo-osmotic shock results in a large difference in the osmotic strength between the intracellular and extracellular spaces. As a result, water rushes into the cell to equalize this gradient increasing the turgor pressure and tension in the cell membrane. If no mechanosensitive channels are present and membrane tension is high (left panel), the membrane ruptures releasing intracellular content into the environment resulting in cell death . If mechanosensitive channels are present (right panel) and membrane tension is beyond the gating tension, the mechanosensitive channel MscL opens, releasing water and small intracellular molecules into the environment thus relieving pressure and membrane tension. (B) The experimental approach undertaken in this work. The number of mechanosensitive channels tagged with a fluorescent reporter is tuned through modification of the Shine-Dalgarno sequence of the *mscL* gene. The cells are then subjected to a hypo-osmotic shock and the number of surviving cells are counted, allowing the calculation of a survival probability.

## Results

### Quantifying the single-cell MscL copy number

The principal goal of this work is to examine the contribution of a single mechanosensitive channel species to cell survival under a hypo-osmotic shock. While this procedure could be performed for any species of channel, we chose MscL as it is the most well characterized and one of the most abundant species in *E. coli*. To probe the contribution of MscL alone, we generated an *E. coli* strain in which all seven known mechanosensitive channel genes were deleted from the chromosome followed by the integration of an *mscL* gene encoding an MscL super-folder GFP (sfGFP) fusion. Chromosomal integration imposes strict control on the gene copy number compared to plasmid borne expression systems, which is important to minimize variation in channel expression across the population and provide conditions more representative of native cell physiology. Fluorescent protein fusions have frequently been used to study MscL and have been shown through electrophysiology to function identically to the native MscL protein, allowing us to confidently draw conclusions about the role this channel plays in wild-type cells from our measurements. (3, 18).

To modulate the number of MscL channels per cell, we developed a series of mutants which were designed to decrease the expression relative to wild-type. These changes involved direct alterations of the Shine-Dalgarno sequence as well as the inclusion of AT hairpins of varying length directly upstream of the start codon which influences the translation rate and hence the number of MscL proteins produced (Fig. 2A). The six Shine-Dalgarno sequences used in this work were chosen using the RBS binding site strength calculator from the Salis Laboratory at the Pennsylvania State University (19, 20). While the designed Shine-Dalgarno sequence mutations decreased the expression relative to wild-type as intended, the distribution of expression is remarkably wide spanning an order of magnitude.

To measure the number of MscL channels per cell, we determined a fluorescence calibration factor to translate arbitrary fluorescence units per cell to protein copy number. While there have been numerous techniques developed over the past decade to directly measure this calibration factor, such as quantifying single-molecule photobleaching constants or measuring the binomial partitioning of fluorescent proteins upon cell division (3, 21), we used *a priori* knowledge of the mean MscL-sfGFP expression level of a particular *E. coli* strain to estimate the average fluorescence of a single channel. In Bialecka-Fornal et al. 2012 (3), the authors used single-molecule photobleaching and quantitative Western blotting to probe the expression of MscL-sfGFP under a wide range of growth conditions. To compute a calibration factor, we used the strain MLG910 (*E. coli* K12 MG1655 φ(mscL-sfGFP)) as a "standard candle", highlighted in yellow in Fig. 2B. This standard candle strain was grown and imaged in identical conditions in which the MscL count was determined. The calibration factor was computed by dividing the mean total cell fluorescence by the known MscL copy number, resulting in a measure of arbitrary fluorescence units per MscL channel. Details regarding this calculation and appropriate propagation of error can be found in the Materials & Methods as well as the supplemental information (*Standard Candle Calibration*).

While it is seemingly trivial to use this calibration to determine the total number of channels per cell for wild-type or highly expressing strains, the calculation for the lowest expressing strains is complicated by distorted cell morphology. We observed that as the channel copy number decreases, cellular morphology becomes increasingly aberrant with filamentous, bulging, and branched cells becoming more abundant (Fig. S3A). This morphological defect has been observed when altering the abundance of several species of mechanosensitive channels, suggesting that they play an important role in general architectural stability (3, 4). As these aberrant morphologies can vary widely in size and shape, calculating the number of channels per cell becomes a more nuanced endeavor. For example, taking the total MscL copy number for these cells could skew the final calculation of survival probability as a large but severely distorted cell would be interpreted as having more channels than a smaller, wild-type shaped cell (Fig. S3B). To correct for this pathology, we computed the average expression level per unit area for each cell and multiplied this by the average cellular area of our standard candle strain which is morphologically indistinguishable from wild-type *E. coli*, allowing for the calculation of an effective channel copy number. The effect of this correction can be seen in Fig. S3C and D, which illustrate that there is no other correlation between cell area and channel expression.

Our calculation of the effective channel copy number for our suite of Shine-Dalgarno mutants is shown in Fig. 2B. The expression of these strains cover nearly three orders of magnitude with the extremes ranging from approximately four channels per cell to nearly one thousand. While the means of each strain are somewhat distinct, the distributions show a large degree of overlap, making one strain nearly indistinguishable from another. This variance is a quantity that is lost in the context of bulk scale experiments but can be accounted for via single-cell methods.

Performing a single-cell hypo-osmotic challenge assay

To measure the channel copy number of a single cell and query its survival after a hypo-osmotic shock, we used a custom-made flow cell in which osmotic shock and growth can be monitored in real time using video microscopy (Fig. 3A). The design and characterization of this device has been described in depth previously and is briefly described in the Materials & Methods (4). Using this device, cells were exposed to a large hypo-osmotic shock by switching between LB Miller medium containing 500mM NaCl and LB media containing no NaCl. All six Shine-Dalgarno modifications shown in Fig. 2B (excluding MLG910) were subjected to a hypo-osmotic shock at controlled rates while under observation. After the application of the osmotic shock, the cells were imaged every sixty seconds for four to six hours. Survivors were defined as cells which underwent at least two divisions post-shock. The brief experimental protocol can be seen in Fig. 3B.

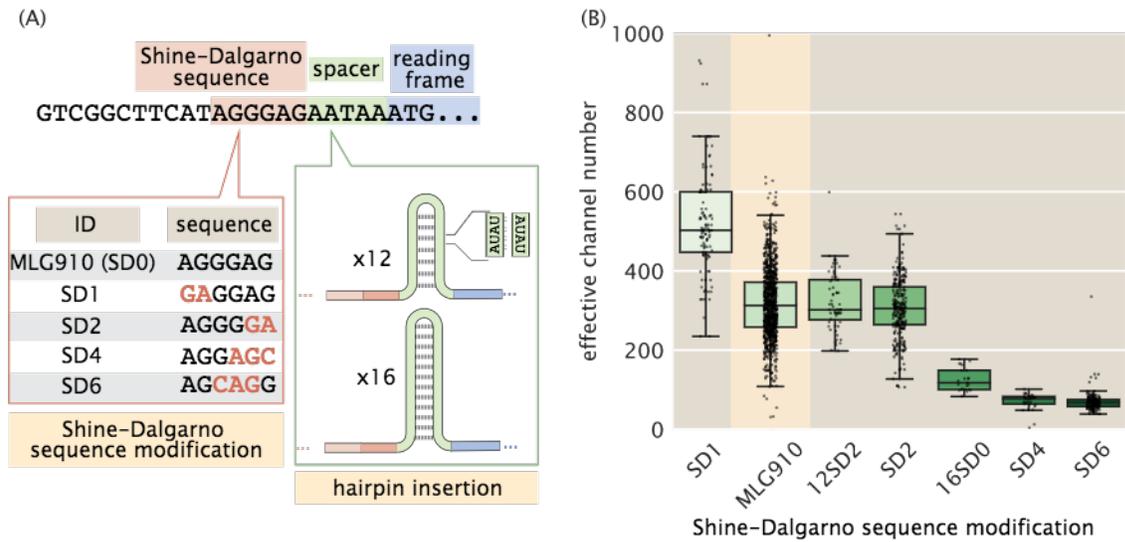

FIG 2 Control of MscL expression and calculation of channel copy number. (A) Schematic view of the expression modifications performed in this work. The beginning portion of the native *mscL* sequence is shown with the Shine-Dalgarno sequence, spacer region, and start codon shaded in red, green, and blue, respectively. The Shine-Dalgarno sequence was modified through the Salis lab Ribosomal Binding Strength calculator (19, 20). The wild-type sequence (MLG910) is shown in black with mutations for the other four Shine-Dalgarno mutants highlighted in red. Expression was further modified by the insertion of repetitive AT bases into the spacer region, generating hairpins of varying length which acted as a thermodynamic barrier for translation initiation. (B) Variability in effective channel copy number is computed using the standard candle. The boxes represent the interquartile region of the distribution, the center line displays the median, and the whiskers represent 1.5 times the maximum and minimum of the interquartile region. Individual measurements are denoted as black points. The strain used for calibration of channel copy number (MLG910) is highlighted in yellow.

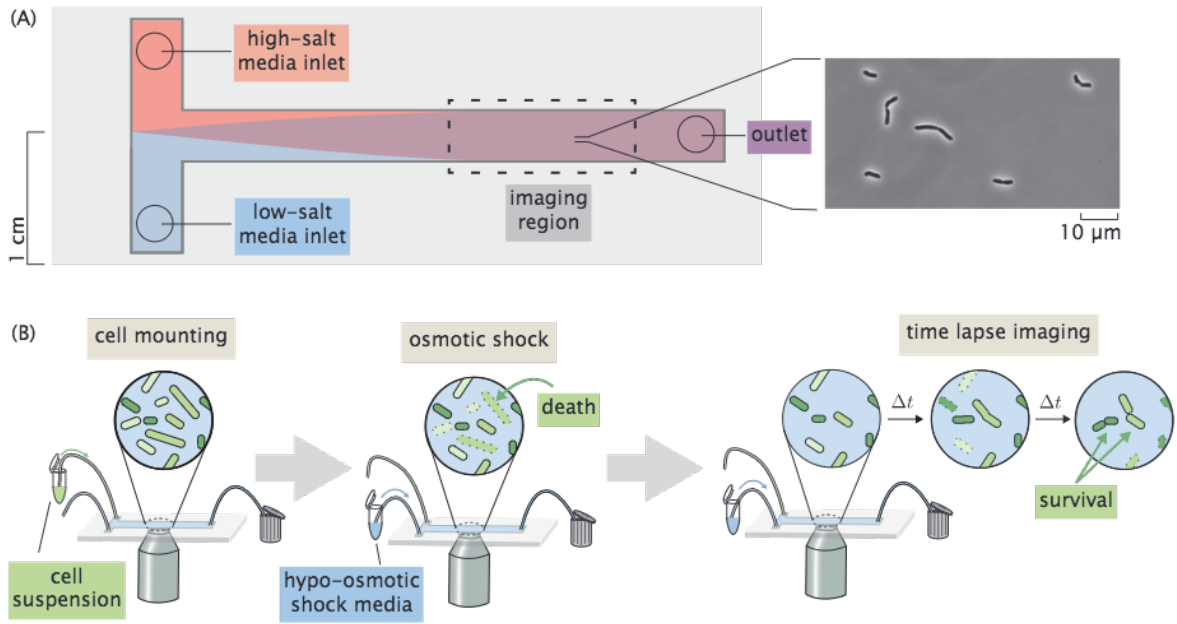

FIG 3 Experimental approach to measuring survival probability. (A) Layout of a home-made flow cell for subjecting cells to osmotic shock. Cells are attached to a polyethylamine functionalized surface of a glass coverslip within the flow chamber by loading a dilute cell suspension through one of the inlets. (B) The typical experimental procedure. Cells are loaded into a flow chamber as shown in (A) and mounted to the glass coverslip surface. Cells are subjected to a hypo-osmotic shock by flowing hypotonic medium into the flow cell. After shock, the cells are monitored for several hours and surviving cells are identified.

Due to the extensive overlap in expression between the different Shine-Dalgarno mutants (see Fig. 2B), computing the survival probability by treating each mutant as an individual bin obfuscates the relationship between channel abundance and survival. To more thoroughly examine this relationship, all measurements were pooled together with each cell being treated as an individual experiment. The hypo-osmotic shock applied in these experiments was varied across a range of 0.02 Hz (complete exchange in 50 s) to 2.2 Hz (complete exchange in 0.45 s). Rather than pooling this wide range of shock rates into a single data set, we chose to separate the data into "slow shock" ( < 1.0 Hz) and "fast shock" ( ≥ 1.0 Hz) classes. Other groupings of shock rate were explored and are discussed in the supplemental information (*Shock Classification*). The cumulative distributions of channel copy number separated by survival are shown in Fig. 4. In these experiments, survival was never observed for a cell containing less than approximately 100 channels per cell, indicated by the red stripe in Fig. 4. This suggests that there is a minimum number of channels needed for survival on the order of 100 per cell. We also observe a slight shift in the surviving fraction of the cells towards higher effective copy number, which matches our intuition that including more mechanosensitive channels increases the survival probability.

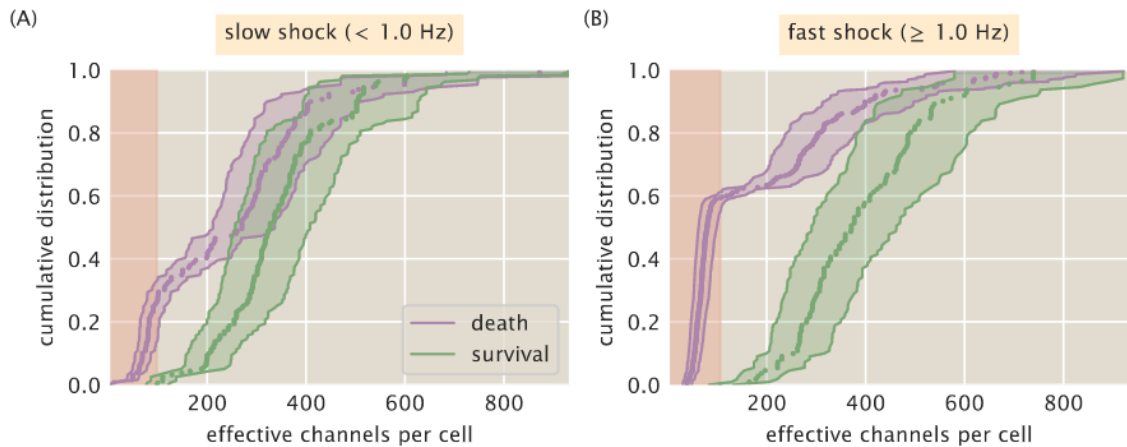

FIG 4 Distributions of survival and death as a function of effective channel number. (A) Empirical cumulative distributions of channel copy number separated by survival (green) or death (purple) after a slow (< 1.0 Hz) osmotic shock. (B) The empirical cumulative distribution for a fast ($\geq$ 1.0 Hz) osmotic shock. Shaded green and purple regions represent the 95% credible region of the effective channel number calculation for each cell. Shaded red stripe signifies the range of channels in which no survival was observed.

Prediction of survival probability as a function of channel copy number

There are several ways by which the survival probability can be calculated. The most obvious approach would be to group each individual Shine-Dalgarno mutant as a single bin and compute the average MscL copy number and the survival probability. Binning by strain is the most frequently used approach for such measurements and has provided valuable insight into the qualitative relationship of survival on other physiological factors (4, 8). However the copy number distribution for each Shine-Dalgarno mutant (Fig. 2B) is remarkably wide and overlaps with the other strains. We argue that this coarse-grained binning negates the benefits of performing single-cell measurements as two strains with different means but overlapping quartiles would be treated as distinctly different distributions.

Another approach would be to pool all data together, irrespective of the Shine-Dalgarno mutation, and bin by a defined range of channels. Depending on the width of the bin, this could allow for finer resolution of the quantitative trend, but the choice of the bin width is arbitrary with the *a priori* knowledge that is available. Drawing a narrow bin width can easily restrict the number of observed events to small numbers where the statistical precision of the survival probability is lost. On the other hand, drawing wide bins increases the precision of the estimate, but becomes further removed from a true single-cell measurement and represents a population mean, even though it may be a smaller population than binning by the Shine-Dalgarno sequence alone. In both of these approaches, it is difficult to extrapolate the quantitative trend outside of the experimentally observed region of channel copy number. Here, we present a method to estimate the probability of survival for any channel copy number, even those that lie outside of the experimentally queried range.

To quantify the survival probability while maintaining single-cell resolution, we chose to use a logistic regression model which does not require grouping data into arbitrary bins and treats each cell measurement as an independent experiment. Logistic regression is an inferential method to model the probability of a boolean or categorical event (such as survival or death) given one or several predictor variables and is commonly used in medical statistics to compute survival rates and dose response curves (22, 23). The primary assumption of logistic regression is that the log-odds probability of survival $p_s$ is linearly dependent on the predictor variable, in our case the log channels per cell $N_c$ with a dimensionless intercept $\beta_0$ and slope $\beta_1$,

$$\log \frac{p_s}{1 - p_s} = \beta_0 + \beta_1 \log N_c. \tag{1}$$

Under this assumption of linearity, $\beta_0$ is the log-odds probability of survival with no MscL channels. The slope $\beta_1$ represents the change in the log-odds probability of survival conveyed by a single channel. As the calculated number of channels in this work spans nearly three orders of magnitude, it is better to perform this regression on $\log N_c$ as regressing on $N_c$ directly would give undue weight for lower channel copy numbers due to the sparse sampling of high-copy number cells. The functional form shown in Eq. 1 can be derived directly from Bayes' theorem and is shown in the supplemental information (*Logistic Regression*). If one knows the values of $\beta_0$ and $\beta_1$, the survival probability can be expressed as

$$p_s = \frac{1}{1 + N_c^{-\beta_1} e^{-\beta_0}}. \tag{2}$$

In this analysis, we used Bayesian inferential methods to determine the most likely values of the coefficients and is described in detail in the supplemental information (*Logistic Regression*).

The results of the logistic regression are shown in Fig. 5. We see a slight rightward shift the survival probability curve under fast shock relative to the slow shock case, reaffirming the conclusion that survival is also dependent on the rate of osmotic shock (4). This rate dependence has been observed for cells expressing MscL alongside other species of mechanosensitive channels, but not for MscL alone. This suggests that MscL responds differently to different rates of shock, highlighting the need for further study of rate dependence and the coordination between different species of mechanosensitive channels. Fig. 5 also shows that several hundred channels are required to provide appreciable protection from osmotic shock. For a survival probability of 80%, a cell must have approximately 500 to 700 channels per cell for a fast and slow shock, respectively. The results from the logistic regression are showed as continuous colored curves. The individual cell measurements separated by survival and death are shown at the top and bottom of each plot, respectively, and are included to provide a sense of sampling density.

Over the explored range of MscL copy number, we observed a maximum of 80% survival for any binning method. The remaining 20% survival may be attained when the other species of mechanosensitive channels are expressed alongside MscL. However, it is possible that the flow cell method performed in this work lowers the maximal survival fraction as the cells are exposed to several, albeit minor, mechanical stresses such as loading into the flow cell and chemical adherence to the glass surface. To ensure that the results from logistic regression accurately describe the data, we can compare the survival probabilities to those using the binning methods described earlier (red and black points, Fig. 5). Nearly all binned data fall within error of the prediction (see Materials & Methods for definition of error bar on probability), suggesting that this approach accurately reflects the survival probability and gives license to extrapolate the estimation of survival probability to regions of outside of our experimentally explored copy number regime.

Thus far, we've dictated that for a given rate of osmotic shock (i.e. "fast" or "slow"), the survival probability is dependent only on the number of channels. In Fig. S7, we show the result of including other predictor variables, such as area and shock rate alone. In such cases, including other predictors resulted in pathological curves showing that channel copy number is the most informative out of the available predictor variables.

## Discussion

One of the most challenging endeavors in the biological sciences is linking the microscopic details of cellular components to the macro-scale physiology of the organism. This formidable task has been met repeatedly in the recent history of biology, especially in the era of DNA sequencing and single molecule biochemistry. For example, the scientific community has been able to connect sickle-cell anemia to a single amino acid substitution in Hemoglobin which promotes precipitation under a change in $O_2$ partial pressure (24–26). Others have assembled a physical model that quantitatively describes chemosensation in bacteria (27) in which the arbiter of sensory adaptation is the repeated methylation of chemoreceptors (28–31). In the past ~50 years alone, numerous biological and physical models of the many facets of the central dogma have been assembled that give us a sense of the interplay between the genome and physiology. For example, the combination of biochemical experimentation and biophysical models have given us a picture of how gene dosage affects furrow positioning in *Drosophila* (32), how recombination of V(D)J gene segments generates an extraordinarily diverse antibody repertoire (33–35), and how telomere shortening through DNA replication is intrinsically tied to cell senescence (36, 37), to name just a few of many such examples.

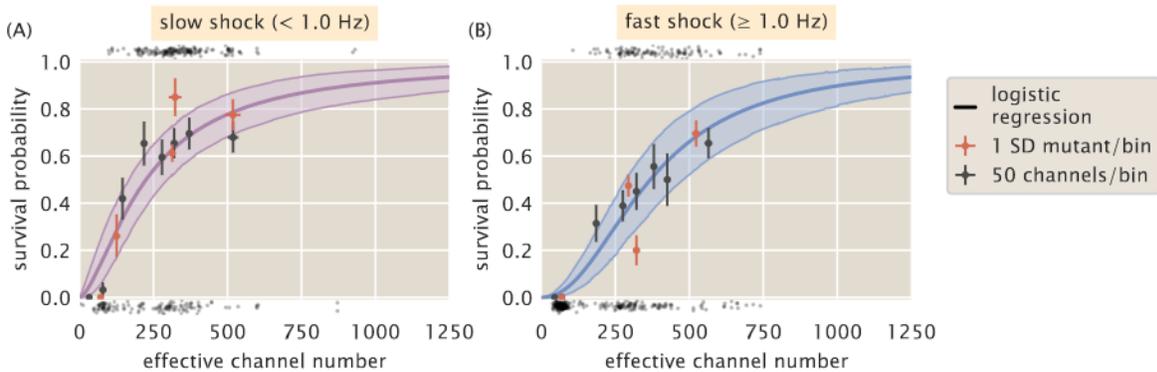

FIG 5 **Probability of survival as a function of MscL copy number.** (A) Estimated survival probability for survival under slow shock as a function of channel copy number. (B) The estimated survival probability of survival under a fast shock as a function of channel copy number. Solid curves correspond to the most probable survival probability from a one-dimensional logistic regression. Shaded regions represent the 95% credible regions. Points at the top and bottom of plots represent individual cell measurements which survived and perished, respectively. The red and black points correspond to the survival probability estimated via binning by Shine-Dalgarno sequence and binning by groups of 50 channels per cell, respectively. Horizontal error bars represent the standard error of the mean from at least 25 measurements. Vertical error bars represent the certainty of the probability estimate given $n$ survival events from $N$ total observations.

By no means are we "finished" with any of these topics. Rather, it's quite the opposite in the sense that having a handle on the biophysical knobs that tune the behavior opens the door to a litany of new scientific questions. In the case of mechanosenstaion and osmoregulation, we have only recently been able to determine some of the basic facts that allow us to approach this fascinating biological phenomenon biophysically. The dependence of survival on mechanosensitive channel abundance is a key quantity that is missing from our collection of critical facts. To our knowledge, this work represents the first attempt to quantitatively control the abundance of a single species of mechanosensitive channel and examine the physiological consequences in terms of survival probability at single-cell resolution. Our results reveal two notable quantities. First, out of the several hundred single-cell measurements, we never observed a cell which had less than approximately 100 channels per cell and survived an osmotic shock, irrespective of the shock rate. The second is that between 500 and 700 channels per cell are needed to provide 80% survival, depending on the shock rate.

Only recently has the relationship between the MscL copy number and the probability of survival been approached experimentally. In van den Berg et al. (2016), the authors examined the contribution of MscL to survival in a genetic background where all other known mechanosensitive channels had been deleted from the chromosome and plasmid-borne expression of an MscL-mEos3.2 fusion was tuned through an IPTG inducible promoter (8). In this work, they measured the single-cell channel abundance through super-resolution microscopy and queried survival through bulk assays. They report a nearly linear relationship between survival and copy number, with approximately 100 channels per cell conveying 100% survival. This number is significantly smaller than our observation of approximately 100 channels as the *minimum* number needed to convey any observable degree of survival.

The disagreement between the numbers reported in this work and in van den Berg et al. may partially arise from subtle differences in the experimental approach. The primary practical difference is the rate and magnitude of the osmotic shock. van den Berg et al. applied an approximately 600 mOsm downshock in bulk at an undetermined rate whereas we applied a 1 Osm downshock at controlled rates varying from 0.02 Hz to 2.2 Hz. In their work, survival was measured through plating assays which represent the population average rather than the distribution of survival probability. While this approach provides valuable information regarding the response of a population to an osmotic shock, the high survival rate may stem from a wide distribution of channel copies per cell in the population coupled with bulk-scale measurement of survival. As has been shown in this work, the expression of MscL from a chromosomal integration is noisy with a single strain exhibiting MscL copy numbers spanning an order of magnitude or more. In van den Berg et al., this variance is exacerbated by expression of MscL from an inducible plasmid as fluctuations in the gene copy number from plasmid replication and segregation influence the expression level. Connecting such a wide and complex distribution of copy numbers to single-cell physiology requires the consideration of moments beyond the mean which is a nontrivial task. Rather than trying to make such a connection, we queried survival at single-cell resolution at the expense of a lower experimental throughput.

Despite these experimental differences, the results of this work and van den Berg et al., are in agreement that MscL must be present at the level of 100 or more channels per cell in wild-type cells to convey appreciable survival. As both of these works were performed in a strain in which the only mechanosensitive channel was MscL, it remains unknown how the presence of the other channel species would alter the number of MscL needed for complete survival. In our experiments, we observed a maximum survival probability of approximately 80% even with close to 1000 MscL channels per cell. It is possible that the combined effort of the six other mechanosensitive channels would make up for some if not all of the remaining 20%. To explore the contribution of another channel to survival, van den Berg et al. also queried the contribution of MscS, another mechanosensitive channel, to survival in the absence of any other species of mechansensitive channel. It was found that over the explored range of MscS channel copy numbers, the maximum survival rate was approximately 50%, suggesting that different mechanosensitive channels have an upper limit to how much protection they can confer. Both van den Berg et al. and our work show that there is still much to be learned with respect to the interplay between the various species of mechanosensitive channel as well as their regulation.

Recent work has shown that both magnitude and the rate of osmotic down shock are important factors in determining cell survival (4). In this work, we show that this finding holds true for a single species of mechanosensitive channel, even at high levels of expression. One might naïvely expect that this rate-dependent effect would disappear once a certain threshold of channels had been met. Our experiments, however, show that even at nearly 1000 channels per cell the predicted survival curves for a slow (< 1.0 Hz) and fast ( 1.0 Hz) are shifted relative to each other with the fast shock predicting lower rates of survival. This suggests either we have not reached this threshold in our experiments or there is more to understand about the relationship between abundance, channel species, and the shock rate.

Some experimental and theoretical treatments suggest that only a few copies of MscL or MscS should be necessary for 100% protection given our knowledge of the conductance and the maximal water flux through the channel in its open state (11, 38). However, recent proteomic studies have revealed average MscL copy numbers to be in the range of several hundred per cell, depending on the condition, as can be seen in Table 1 (15, 16, 39). Studies focusing solely on MscL have shown similar counts through quantitative Western blotting and fluorescence microscopy (3). Electrophysiology studies have told another story with copy number estimates ranging between 4 and 100 channels per cell (17, 40). These measurements, however, measure the active number of channels. The factors regulating channel activity in these experiments could be due to perturbations during the sample preparation or reflect some unknown mechanism of regulation, such as the presence or absence of interacting cofactors (41). The work described here, on the other hand, measures the *maximum* number of channels that could be active and may be able to explain why the channel abundance is higher than estimated by theoretical means. There remains much more to be leared about the regulation of activity in these systems. As the *in vivo* measurement of protein copy number becomes accessible through novel single-cell and single-molecule methods, we will continue to collect more facts about this fascinating system and hopefully connect the molecular details of mechanosensation with perhaps the most important physiological response – life or death.

TABLE 1 Measured cellular copy numbers of MscL. Asterisk (*) Indicates inferred MscL channel copy number from the total number of detected MscL peptides.

| Reported channels per cell | Method | Reference |
| --- | --- | --- |
| 480 ± 103 | Western blotting | (3) |
| 560* | Ribosomal profiling | (39) |
| 331* | Mass spectrometry | (15) |
| 583* | Mass spectrometry | (16) |
| 4 - 5 | Electrophysiology | (17) |
| 10 - 100 | Electrophysiology | (13) |
| 10 - 15 | Electrophysiology | (40) |

## Materials & Methods

### Bacterial strains and growth conditions

The bacterial strains are described in Table S1. The parent strain for the mutants used in this study was MJF641 (5), a strain which had all seven mechanosensitive channels deleted. The MscL-sfGFP coding region from MLG910 (3) was integrated into MJF641 by P1 transduction, creating the strain D6LG-Tn10. Selection pressure for MscL integration was created by incorporating an osmotic shock into the transduction protocol, which favored the survival of MscL-expressing stains relative to MJF641 by ~100-fold. Screening for integration candidates was based on fluorescence expression of plated colonies. Successful integration was verified by sequencing. Attempts to transduce RBS-modified MscL-sfGFP coding regions became increasingly inefficient as the targeted expression level of MscL was reduced. This was due to the decreasing fluorescence levels and survival rates of the integration candidates. Consequently, Shine-Dalgarno sequence modifications were made by inserting DNA oligos with lambda Red-mediated homologous recombination, i.e., recombineering (42). The oligos had a designed mutation (Fig. 2) flanked by ~25 base pairs that matched the targeted MscL region (Table S2). A two-step recombineering process of selection followed by counter selection using a *tetA-sacB* gene fusion cassette (43) was chosen because of its capabilities to integrate with efficiencies comparable to P1 transduction and not leave antibiotic resistance markers or scar sequences in the final strain. To prepare the strain D6LG-Tn10 for this scheme, the Tn10 transposon containing the *tetA* gene needed to be removed to avoid interference with the *tetA-sacB* cassette. Tn10 was removed from the middle of the *ycjM* gene with the primer Tn10delR (Table S2) by recombineering, creating the strain D6LG (SD0). Counter selection against the *tetA* gene was promoted by using agar media with fusaric acid (43, 44). The *tetA-sacB* cassette was PCR amplified out of the strain XTL298 using primers MscLSPSac and MscLSPSacR (Table S2). The cassette was integrated in place of the spacer region in front of the MscL start codon of D6LG (SD0) by recombineering, creating the intermediate strain D6LTetSac. Positive selection for cassette integration was provided by agar media with tetracycline. Finally, the RBS modifying oligos were integrated into place by replacing the *tetA-sacB* cassette by recombineering. Counter selection against both *tetA* and *sacB* was ensured by using agar media with fusaric acid and sucrose (43), creating the Shine-Dalgarno mutant strains used in this work.

Strain cultures were grown in 5 mL of LB-Lennox media with antibiotic (apramycin) overnight at 37°C. The next day, 50 µL of overnight culture was inoculated into 5 mL of LB-Lenox with antibiotic and the culture was grown to OD600nm ~0.25. Subsequently, 500 µL of that culture was inoculated into 5 mL of LB-Lennox supplemented with 500mM of NaCl and the culture was regrown to OD600nm ~0.25. A 1 mL aliquot was taken and used to load the flow cell.

Flow cell

All experiments were conducted in a home-made flow cell as is shown in Fig. 3A. This flow cell has two inlets which allow media of different osmolarity to be exchanged over the course of the experiment. The imaging region is approximately 10 mm wide and 100 $\mu$m in depth. All imaging took place within 1 – 2 cm of the outlet to avoid imaging cells within a non-uniform gradient of osmolarity. The interior of the flow cell was functionalized with a 1:400 dilution of polyethylamine prior to addition of cells with the excess washed away with water. A dilute cell suspension in LB Lennox with 500 mM NaCl was loaded into one inlet while the other was connected to a vial of LB medium with no NaCl. This hypotonic medium was clamped during the loading of the cells.

Once the cells had adhered to the polyethylamine coated surface, the excess cells were washed away with the 500 mM NaCl growth medium followed by a small (~20 $\mu$L) air bubble. This air bubble forced the cells to lay flat against the imaging surface, improving the time-lapse imaging. Over the observation period, cells not exposed to an osmotic shock were able to grow for 4 – 6 divisions, showing that the flow cell does not directly impede cell growth.

Imaging conditions

All imaging was performed in a flow cell held at 30°C on a Nikon Ti-Eclipse microscope outfitted with a Perfect Focus system enclosed in a Haison environmental chamber (approximately 1°C regulation efficiency). The microscope was equipped with a 488 nm laser excitation source (CrystaLaser) and a 520/35 laser optimized filter set (Semrock). The images were collected on an Andor Xion +897 EMCCD camera and all microscope and acquisition operations were controlled via the open source µManager microscope control software (27). Once cells were securely mounted onto the surface of the glass coverslip, between 15 and 20 positions containing 5 to 10 cells were marked and the coordinates recorded. At each position, a phase contrast and GFP fluorescence image was acquired for segmentation and subsequent measurement of channel copy number. To perform the osmotic shock, LB media containing no NaCl was pulled into the flow cell through a syringe pump. To monitor the media exchange, both the high salt and no salt LB media were supplemented with a low-affinity version of the calcium-sensitive dye Rhod-2 (250 nM; TEF Labs) which fluoresces when bound to $Ca^{2+}$. The no salt medium was also supplemented with 1µM $CaCl_2$ to make the media mildly fluorescent and the exchange rate was calculated by measuring the fluorescence increase across an illuminated section of one of the positions. These images were collected in real time for the duration of the shock. The difference in measured fluorescence between the pre-shock images and those at the end of the shock set the scale of a 500 mM NaCl down shock. The rate was calculated by fitting a line to the middle region of this trace. Further details regarding this procedure can be found in Bialecka-Fornal, Lee, and Phillips, 2015 (4).

Image Processing

Images were processed using a combination of automated and manual methods. First, expression of MscL was measured via segmenting individual cells or small clusters of cells in phase contrast and computing the mean pixel value of the fluorescence image for each segmented object. The fluorescence images were passed through several filtering operations which reduced high-frequency noise as well as corrected for uneven illumination of the excitation wavelength.

Survival or death classification was performed manually using the CellProfiler plugin for ImageJ software (NIH). A survivor was defined as a cell which was able to undergo two division events after the osmotic down shock. Cells which detached from the surface during the post-shock growth phase or those which became indistinguishable from other cells due to clustering were not counted as survival or death and were removed from the dataset completely. A region of the cell was manually marked with 1.0 (survival) or 0.0 (death) by clicking on the image. The `xy` coordinates of the click as well as the assigned value were saved as an `.xml` file for that position.

The connection between the segmented cells and their corresponding manual markers was automated. As the manual markings were made on the first phase contrast image after the osmotic shock, small shifts in the positions of the cell made one-to-one mapping with the segmentation mask non-trivial. The linkages between segmented cell and manual marker were made by computing all pairwise distances between the manual marker and the segmented cell centroid, taking the shortest distance as the true pairing. The linkages were then inspected manually and incorrect mappings were corrected as necessary.

All relevant statistics about the segmented objects as well as the sample identity, date of acquisition, osmotic shock rate, and camera exposure time were saved as `.csv` files for each individual experiment. A more in-depth description of the segmentation procedure as well as the relevant code can be accessed as a Jupyter Notebook at (`http://rpgroup.caltech.edu/mscl_survival`).

Calculation of effective channel copy number

To compute the MscL channel copy number, we relied on measuring the fluorescence level of a bacterial strain in which the mean MscL channel copy number was known via fluorescence microscopy (3). *E. coli* strain MLG910, which expresses the MscL-sfGFP fusion protein from the wild-type SD sequence, was grown under identical conditions to those described in Bialecka-Fornal et al. 2015 in M9 minimal medium supplemented with 0.5% glucose to an OD$_{600nm}$ of ~0.3. The cells were then diluted ten fold and immobilized on a rigid 2% agarose substrate and placed onto a glass bottom petri dish and imaged in the same conditions as described previously.

Images were taken of six biological replicates of MLG910 and were processed identically to those in the osmotic shock experiments. A calibration factor between the average cell fluorescence level and mean MscL copy number was then computed. We assumed that all measured fluorescence (after filtering and background subtraction) was derived from the MscL-sfGFP fusion,

$$I_{\text{tot}} = \alpha N, \qquad (3)$$

in which $\alpha$ is the calibration factor and $N$ is the mean cellular MscL-sfGFP copy number as reported in Bialecka-Fornal et al. 2012 (3). To correct for errors in segmentation, the intensity was computed as an areal density $I_A$ and was multiplied by the average cell area $A$ of the population. The calibration factor was therefore computed as

$$\alpha = \frac{I_A \, A}{N}. \tag{4}$$

We used Bayesian inferential methods to compute this calibration factor taking measurement error and replicate-to-replicate variation into account. The resulting average cell area and calibration factor was used to convert the measured cell intensities from the osmotic shock experiments to cell copy number. The details of this inference are described in depth in the supplemental information (*Standard Candle Calibration*).

Logistic regression

We used Bayesian inferential methods to find the most probable values of the coefficients $\beta_0$ and $\beta_1$ and the appropriate credible regions and is described in detail in the supplemental information (*Logistic Regression*). Briefly, we used Markov chain Monte Carlo (MCMC) to sample from the log posterior distribution and took the most probable value as the mean of the samples for each parameter. The MCMC was performed using the Stan probabilistic programming language (45) and all models can be found on the GitHub repository (`http://github.com/rpgroup-pboc/mscl_survival`).

Calculation of survival probability error

The vertical error bars for the points shown in Fig. 5 represent our uncertainty in the survival probability given our measurement of $n$ survivors out of a total $N$ single-cell measurements. The probability distribution of the survival probability $p_s$ given these measurements can be written using Bayes' theorem as

$$g(p_s \mid n, N) = \frac{f(n \mid p_s, N) g(p_s)}{f(n \mid N)}, \tag{5}$$

where $g$ and $f$ represent probability density functions over parameters and data, respectively. The likelihood $f(n \mid p_s, N)$ represents the probability of measuring $n$ survival events, given a total of $N$ measurements each with a probability of survival $p_s$. This matches the story for the Binomial distribution and can be written as

$$f(n \mid p_s, N) = \frac{N!}{n!(N-n)!} p_s^n (1-p_s)^{N-n}. \tag{6}$$

To maintain maximal ignorance we can assume that any value for $p_s$ is valid, such that is in the range [0, 1]. This prior knowledge, represented by $g(p_s)$, can be written as

$$g(p_s) = \begin{cases} 1 & 0 \le p_s \le 1 \\ 0 & \text{otherwise} \end{cases}. \tag{7}$$

We can also assume maximal ignorance for the total number of survival events we could measure given $N$ observations, $f(n\,|\,N)$. Assuming all observations are equally likely, this can be written as

$$f(n\,|\,N) = \frac{1}{N+1} \tag{8}$$

where the addition of one comes from the possibility of observing zero survival events. Combining Eqns. 6, 7, 8, the posterior distribution $g(p_s\,|\,n,N)$ is

$$g(p_s\,|\,n,N) = \frac{(N+1)!}{n!(N-n)!} p_s^n (1-p_s)^{N-n}. \tag{9}$$

The most probable value of $p_s$, where the posterior probability distribution given by Eq. 9 is maximized, can be found by computing the point at which derivative of the log posterior with respect to $p_s$ goes to zero,

$$\frac{d \log g(p_s\,|\,n,N)}{dp_s} = \frac{n}{p_s} - \frac{N-n}{1-p_s} = 0. \tag{10}$$

Solving Eq. 10 for $p_s$ gives the most likely value for the probability,

$$p_s = \frac{n}{N}. \tag{11}$$

So long as $N \gg np_s$, Eq. 9 can be approximated as a Gaussian distribution with a mean $p_s$ and a variance $\sigma_{p_s}^2$. By definition, the variance of a Gaussian distribution is computed as the negative reciprocal of the second derivative of the log posterior evaluated at $p_s = p_s$,

$$\sigma_{p_s}^2 = -\left[ \frac{d^2 \log g(p_s\,|\,n,N)}{dp_s^2} \bigg|_{p_s = p_s} \right]^{-1}. \tag{12}$$

Evaluating Eq. 12 yields

$$\sigma_{p_s}^2 = \frac{n(N-n)}{N^3}. \tag{13}$$

Given Eq. 11 and Eq. 13, the most-likely survival probability and estimate of the uncertainty can be expressed as

$$p_s = p_s \pm \sigma_{p_s}. \tag{14}$$

### Data and software availability

All raw image data is freely available and is stored on the CaltechDATA Research Data Repository (46). The raw Markov chain Monte Carlo samples are stored as .csv files on CaltechDATA (47). All processed experimental data, Python, and Stan code used in this work are freely available through our GitHub repository (http://github.com/rpgroup-pboc/mscl_survival)(48) accessible through DOI: 10.5281/zenodo.1252524. The scientific community is invited to fork our repository and open constructive issues.

Acknowledgements

We thank Nathan Belliveau, Maja Bialecka-Fornal, Justin Bois, Soichi Hirokawa, Jaspar Landman, Manuel Razo-Mejia, Muir Morrison, and Shyam Saladi for useful advice and discussion. We thank Don Court for strain XTL298 and Samantha Miller for strain MJF641. This work was supported by the National Institutes of Health DP1 OD000217 (Director's Pioneer Award), R01 GM085286, GM084211-A1 , GM118043-01, and La Fondation Pierre Gilles de Gennes.

# Supplemental information for "Connecting the dots between mechanosensitive channel abundance, osmotic shock, and survival at single-cell resolution"

Griffin Chure$^{a,\dagger}$, Heun Jin Lee$^{b,\dagger}$, and Rob Phillips$^{a,b,}$

$^a$Department of Biology and Biological Engineering, California Insitute of Technology, Pasadena, CA, USA

$^b$Department of Physics, California Institute of Technology, Pasadena, CA, USA

† contributed equally

* Send correspondence to `phillips@pboc.caltech.edu`

## Standard Candle Calibration

To estimate the single-cell MscL abundance via microscopy, we needed to determine a calibration factor that could translate arbitrary fluorescence units to protein copy number. To compute this calibration factor, we relied on *a priori* knowledge of the mean copy number of MscL-sfGFP for a particular bacterial strain in specific growth conditions. In Bialecka-Fornal et al. 2012 (3), the average MscL copy number for a population of cells expressing an MscL-sfGFP fusion (*E. coli* K-12 MG1655 φ(*mscL-sfGFP*)) cells was measured using quantitative Western blotting and single-molecule photobleaching assays. By growing this strain in identical growth and imaging conditions, we can make an approximate measure of this calibration factor. In this section, we derive a statistical model for estimating the most-likely value of this calibration factor and its associated error.

## Definition of a calibration factor

We assume that all detected fluorescence signal from a particular cell is derived from the MscL-sfGFP protein, after background subtraction and correction for autofluorescence. The arbitrary units of fluorescence can be directly related to the protein copy number via a calibration factor $\alpha$,

$$I_{tot} = \alpha N_{tot}, \tag{15}$$

where $I_{tot}$ is the total cell fluorescence and $N_{tot}$ is the total number of MscL proteins per cell. Bialecka-Fornal et al. report the average cell MscL copy number for the population rather than the distribution. Knowing only the mean, we can rewrite Eq. 15 as

$$\langle I_{tot} \rangle = \alpha \langle N_{tot} \rangle, \tag{16}$$

assuming that $\alpha$ is a constant value that does not change from cell to cell or fluorophore to fluorophore.

The experiments presented in this work were performed using non-synchronously growing cultures. As there is a uniform distribution of growth phases in the culture, the cell size distribution is broad the the extremes being small, newborn cells and large cells in the process of division. As described in the main text, the cell size distribution of a population is broadened further by modulating the MscL copy number with low copy numbers resulting in aberrant cell morphology. To speak in the terms of an effective channel copy number, we relate the average areal intensity of the population to the average cell size,

$$\langle I_{tot} \rangle = \langle I_A \rangle \langle A \rangle = \alpha \langle N_{tot} \rangle, \tag{17}$$

where $\langle I_A \rangle$ is the average areal intensity in arbitrary units per pixel of the population and $\langle A \rangle$ is the average area of a segmented cell. As only one focal plane was imaged in these experiments, we could not compute an appropriate volume for each cell given the highly aberrant morphology. We therefore opted to use the projected two-dimensional area of each cell as a proxy for cell size. Given this set of measurements, the calibration factor can be computed as

$$\alpha = \frac{\langle I_A \rangle \langle A \rangle}{\langle N_{tot} \rangle}. \tag{18}$$

While it is tempting to use Eq. 18 directly, there are multiple sources of error that are important to propagate through the final calculation. The most obvious error to include is the measurement error reported in Bialecka-Fornal et al. 2012 for the average MscL channel count (3). There are also slight variations in expression across biological replicates that arise from a myriad of day-to-day differences. Rather than abstracting all sources of error away into a systematic error budget, we used an inferential model derived from Bayes' theorem that allows for the computation of the probability distribution of $\alpha$.

## Estimation of α for a single biological replicate

A single data set consists of several hundred single-cell measurements of intensity, area of the segmentation mask, and other morphological quantities. The areal density $I_A$ is computed by dividing the total cell fluorescence by the cell area $A$. We are interested in computing the probability distributions for the calibration factor α, the average cell area $\langle A \rangle$, and the mean number of channels per cell $N_{\text{tot}}$ for the data set as a whole given only $I_A$ and $A$. Using Bayes' theorem, the probability distribution for these parameters given a single cell measurement, hereafter called the posterior distribution, can be written as

$$g(\alpha, \langle A \rangle, N_{\text{tot}} \mid A, I_A) = \frac{f(A, I_A \mid \alpha, \langle A \rangle, N_{\text{tot}}) g(\alpha, \langle A \rangle, N_{\text{tot}})}{f(\alpha, I_A)}, \tag{19}$$

where $g$ and $f$ represent probability density functions over parameters and data, respectively. The term $f(A, I_A \mid \alpha, \langle A \rangle, N_{\text{tot}})$ in the numerator represents the likelihood of observing the areal intensity $I_A$ and area $A$ of a cell for a given values of α, $\langle A \rangle$, and $N_{\text{tot}}$. The second term in the numerator $g(\alpha, \langle A \rangle, N_{\text{tot}})$ captures all prior knowledge we have regarding the possible values of these parameters knowing nothing about the measured data. The denominator, $f(I_A, A)$ captures the probability of observing the data knowing nothing about the parameter values. This term, in our case, serves simply as a normalization constant and is neglected for the remainder of this section.

To determine the appropriate functional form for the likelihood and prior, we must make some assumptions regarding the biological processes that generate them. As there are many independent processes that regulate the timing of cell division and cell growth, such as DNA replication and peptidoglycan synthesis, it is reasonable to assume that for a given culture the distribution of cell size would be normally distributed with a mean of $\langle A \rangle$ and a variance $\sigma_{\langle A \rangle}$. Mathematically, we can write this as

$$f(A \mid \langle A \rangle, \sigma_{\langle A \rangle}) \propto \frac{1}{\sigma_{\langle A \rangle}} \exp\left[-\frac{(A - \langle A \rangle)^2}{2\sigma_{\langle A \rangle}^2}\right], \tag{20}$$

where the proportionality results from dropping normalization constants for notational simplicity.

While total cell intensity is intrinsically dependent on the cell area the areal intensity $I_A$ is independent of cell size. The myriad processes leading to the detected fluorescence, such as translation and proper protein folding, are largely independent, allowing us to assume a normal distribution for $I_A$ as well with a mean $\langle I_A \rangle$ and a variance $\sigma_{I_A}^2$. However, we do not have knowledge of the average areal intensity for the standard candle strain *a priori*. This can be calculated knowing the calibration factor, total MscL channel copy number, and the average cell area as

$$\langle I_A \rangle = \frac{\alpha \langle N_{\text{tot}} \rangle}{\langle A \rangle}. \tag{21}$$

Using Eq. 21 to calculate the expected areal intensity for the population, we can write the likelihood as a

Gaussian distribution,

$$f(I_A \mid \alpha, A, N_{\text{tot}}, \sigma_{I_A}) \propto \frac{1}{\sigma_{I_A}} \exp\left[-\frac{\left(I_A - \frac{\alpha N_{\text{tot}}}{A}\right)^2}{2\sigma_{I_A}^2}\right]. \tag{22}$$

With these two likelihoods in hand, we are tasked with determining the appropriate priors. As we have assumed normal distributions for the likelihoods of $A$ and $I_A$, we have included two additional parameters, $\sigma_A$ and $\sigma_{I_A}$, each requiring their own prior probability distribution. It is common practice to assume maximum ignorance for these variances and use a Jeffreys prior (49),

$$g(\sigma_A, \sigma_{I_A}) = \frac{1}{\sigma_A \sigma_{I_A}}. \tag{23}$$

The next obvious prior to consider is for the average channel copy number $N_{\text{tot}}$, which comes from Bialecka-Fornal et al. 2012. In this work, they report a mean $\mu_N$ and variance $\sigma_N^2$, allowing us to assume a normal distribution for the prior,

$$g(N_{\text{tot}} \mid \mu_N, \sigma_N) \propto \frac{1}{\sigma_N} \exp\left[-\frac{(N_{\text{tot}} - \mu_N)^2}{2\sigma_N^2}\right]. \tag{24}$$

For $\alpha$ and $A$, we have some knowledge of what these parameters can and cannot be. For example, we know that neither of these parameters can be negative. As we have been careful to not overexpose the microscopy images, we can say that the maximum value of $\alpha$ would be the bit-depth of our camera. Similarly, it is impossible to segment a single cell with an area larger than our camera's field of view (although there are biological limitations to size below this extreme). To remain maximally uninformative, we can assume that the parameter values are uniformly distributed between these bounds, allowing us to state

$$g(\alpha) = \begin{cases} \frac{1}{\alpha_{\max} - \alpha_{\min}} & \alpha_{\min} \leq \alpha \leq \alpha_{\max} \\ 0 & \text{otherwise} \end{cases}, \tag{25}$$

for $\alpha$ and

$$g(A) = \begin{cases} \frac{1}{A_{\max} - A_{\min}} & A_{\min} \leq A \leq A_{\max} \\ 0 & \text{otherwise} \end{cases} \tag{26}$$

for $A$.

Piecing Eq. 20 through Eq. 26 together generates a complete posterior probability distribution for the parameters given a single cell measurement. This can be generalized to a set of $k$ single cell

measurements as

$$g(\alpha, \bar{A}, N_{\text{tot}}, \sigma_{I_A}, \sigma_A \mid [I_A, A], \mu_N, \sigma_N) \propto \frac{1}{(\alpha_{\max} - \alpha_{\min})(\bar{A}_{\max} - \bar{A}_{\min})} \frac{1}{(\sigma_{I_A} \sigma_A)^{k+1}} \times$$

$$\frac{1}{\sigma_N} \exp\left[-\frac{(N_{\text{tot}} - \mu_N)^2}{2\sigma_N^2}\right] \prod_i^k \exp\left[-\frac{(A^{(i)} - \bar{A})^2}{2\sigma_A^2} - \frac{(I_A^{(i)} - \frac{\alpha N_{\text{tot}}}{\bar{A}})^2}{2\sigma_{I_A}^2}\right],$$

(27)

where $[I_A, A]$ represents the set of $k$ single-cell measurements.

As small variations in the day-to-day details of cell growth and sample preparation can alter the final channel count of the standard candle strain, it is imperative to perform more than a single biological replicate. However, properly propagating the error across replicates is non trivial. One option would be to pool together all measurements of $n$ biological replicates and evaluate the posterior given in Eq. 27. However, this by definition assumes that there is no difference between replicates. Another option would be to perform this analysis on each biological replicate individually and then compute a mean and standard deviation of the resulting most-likely parameter estimates for $\alpha$ and $\bar{A}$. While this is a better approach than simply pooling all data together, it suffers a bias from giving each replicate equal weight, skewing the estimate of the most-likely parameter value if one replicate is markedly brighter or dimmer than the others. Given this type of data and a limited number of biological replicates ($n = 6$ in this work), we chose to extend the Bayesian analysis presented in this section to model the posterior probability distribution for $\alpha$ and $\bar{A}$ as a hierarchical process in which $\alpha$ and $\bar{A}$ for each replicate is drawn from the same distribution.

### A hierarchical model for estimating α

In the previous section, we assumed maximally uninformative priors for the most-likely values of $\alpha$ and $\bar{A}$. While this is a fair approach to take, we are not completely ignorant with regard to how these values are distributed across biological replicates. A major assumption of our model is that the most-likely value of $\alpha$ and $\bar{A}$ for each biological replicate are comparable, so long as the experimental error between them is minimized. In other words, we are assuming that the most-likely value for each parameter for each replicate is drawn from the same distribution. While each replicate may have a unique value, they are all related to one another. Unfortunately, proper sampling of this distribution requires an extensive amount of experimental work, making inferential approaches more attractive.

This approach, often called a multi-level or hierarchical model, is schematized in Fig. 6. Here, we use an informative prior for $\alpha$ and $\bar{A}$ for each biological replicate. This informative prior probability distribution can be described by summary statistics, often called hyper-parameters, which are then treated as the "true" value and are used to calculate the channel copy number. As this approach allows us to get a picture of the probability distribution of the hyper-parameters, we are able to report a point estimate for the most-likely value along with an error estimate that captures all known sources of variation.

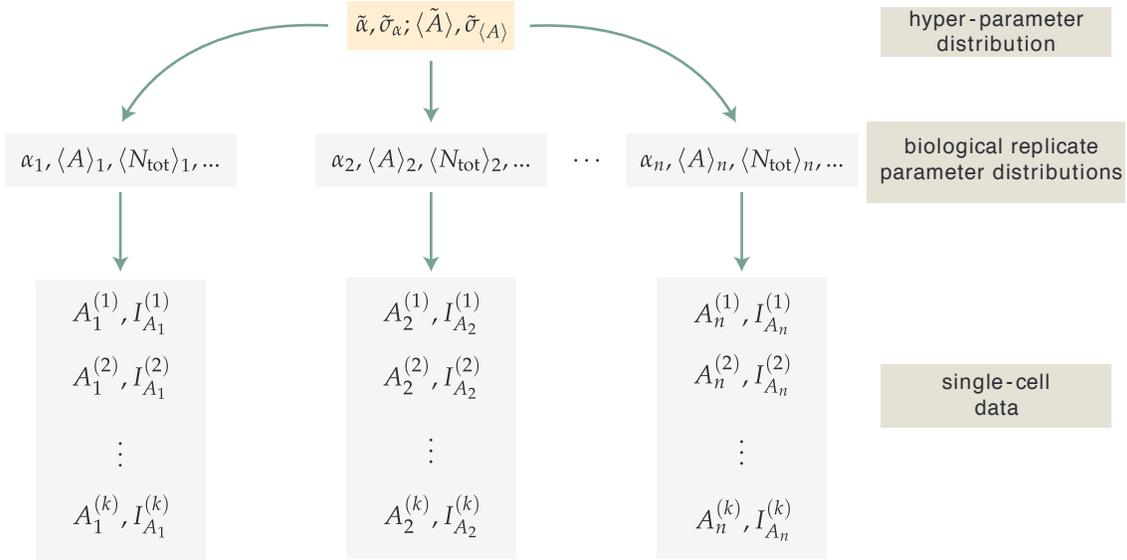

FIG 6 Schematic of hierarchical model structure. The hyper-parameter probability distributions (top panel) are used as an informative prior for the most-likely parameter values for each biological replicate (middle panel). The single-cell measurements of cell area and areal intensity (bottom panel) are used as data in the evaluation of the likelihood.

The choice for the functional form for the informative prior is often not obvious and can require other experimental approaches or back-of-the-envelope estimates to approximate. Each experiment in this work was carefully constructed to minimize the day-to-day variation. This involved adhering to well-controlled growth temperatures and media composition, harvesting of cells at comparable optical densities, and ensuring identical imaging parameters. As the experimental variation is minimized, we can use our knowledge of the underlying biological processes to guess at the approximate functional form. For similar reasons presented in the previous section, cell size is controlled by a myriad of independent processes. As each replicate is independent of another, it is reasonable to assume a normal distribution for the average cell area for each replicate. This normal distribution is described by a mean $\tilde{A}$ and variance $\tilde{\sigma}_A$. Therefore, the prior for $A$ for $n$ biological replicates can be written as

$$g(A \mid \tilde{A}, \tilde{\sigma}_A) \propto \frac{1}{\tilde{\sigma}_A^n} \prod_{j=1}^{n} \exp\left[-\frac{(A_j - \tilde{A})^2}{2\tilde{\sigma}_A^2}\right]. \qquad (28)$$

In a similar manner, we can assume that the calibration factor for each replicate is normally distributed with a mean $\tilde{\alpha}$ and variance $\tilde{\sigma}_\alpha$,

$$g(\alpha \mid \tilde{\alpha}, \tilde{\sigma}_\alpha) \propto \frac{1}{\tilde{\sigma}_\alpha^n} \prod_{j=1}^{n} \exp\left[-\frac{(\alpha_j - \tilde{\alpha})^2}{2\tilde{\sigma}_\alpha^2}\right]. \tag{29}$$

With the inclusion of two more normal distributions, we have introduced four new parameters, each of which needing their own prior. However, our knowledge of the reasonable values for the hyper-parameters has not changed from those described for a single replicate. We can therefore use the same maximally uninformative Jeffreys priors given in Eq. 23 for the variances and the uniform distributions given in Eq. 25 and Eq. 26 for the means. Stitching all of this work together generates the full posterior probability distribution for the best-estimate of $\tilde{\alpha}$ and $\tilde{A}$ shown in Eq. 16 given $n$ replicates of $k$ single cell measurements,

$$g(\tilde{\alpha}, \tilde{\sigma}_\alpha, \tilde{A}, \tilde{\sigma}_A, \{N_{\text{tot}}, A, \alpha, \sigma_{I_A}\} \mid [I_A, A], \mu_N, \sigma_N) \propto$$

$$\frac{1}{(\tilde{\alpha}_{\max} - \tilde{\alpha}_{\min})(\tilde{A}_{\max} - \tilde{A}_{\min})\sigma_N^n (\tilde{\sigma}_\alpha \tilde{\sigma}_A)^{n+1}} \times$$

$$\prod_{j=1}^{n} \exp\left[-\frac{(N_j^{(i)} - \mu_N)^2}{2\sigma_N^2} - \frac{(\alpha_j - \tilde{\alpha})^2}{2\tilde{\sigma}_\alpha^2} - \frac{(A_j - \tilde{A})^2}{2\tilde{\sigma}_A^2}\right] \times$$

$$\frac{1}{(\sigma_{I_{Aj}} \sigma_{A_j})^{k_j+1}} \prod_{i=1}^{k_j} \exp\left[-\frac{(A_j^{(i)} - A_j)^2}{2\sigma_{A_j}^{(i)2}} - \frac{\left(I_{Aj}^{(i)} - \frac{\alpha_j N_{\text{tot}\,j}}{A_j}\right)^2}{2\sigma_{I_{Aj}}^{(i)2}}\right], \tag{30}$$

where the braces $\{\ldots\}$ represent the set of parameters for biological replicates and the brackets $[\ldots]$ correspond to the set of single-cell measurements for a given replicate.

While Eq. 30 is not analytically solvable, it can be easily sampled using Markov chain Monte Carlo (MCMC). The results of the MCMC sampling for $\tilde{\alpha}$ and $\tilde{A}$ can be seen in Fig. 7. From this approach, we found the most-likely parameter values of $3300^{+700}_{-700}$ a.u. per MscL channel and $5.4^{+0.4}_{-0.5}$ $\mu$m$^2$ for $\tilde{\alpha}$ and $\tilde{A}$, respectively. Here, we've reported the median value of the posterior distribution for each parameter with the upper and lower bound of the 95% credible region as superscript and subscript, respectively. These values and associated errors were used in the calculation of channel copy number.

### Effect of correction

The posterior distributions for $\alpha$ and $A$ shown in Fig. 7 were used directly to compute the most-likely channel copy number for each measurement of the Shine-Dalgarno mutant strains, as is described in the coming section (*Logistic Regression*). The importance of this correction can be seen in Fig. 8. Cells with low abundance of MscL channels exhibit notable morphological defects, as illustrated in Fig. 8A. While these would all be considered single cells, the two-dimensional area of each may be comparable to two or three wild-type cells. For all of the Shine-Dalgarno mutants, the distribution of projected cell area has a long tail, with the extremes reaching 35 µm² per cell (Fig. 8B). Calcuating the total number of channels per cell does nothing to decouple this correlation between cell area and measured cell intensity. Fig. 8C shows the correlation between cell area and the total number of channels without normalizing to an average cell size $A$ differentiated by their survival after an osmotic down-shock. This correlation is removed by calculating an effective channel copy number shown in Fig. 8D.

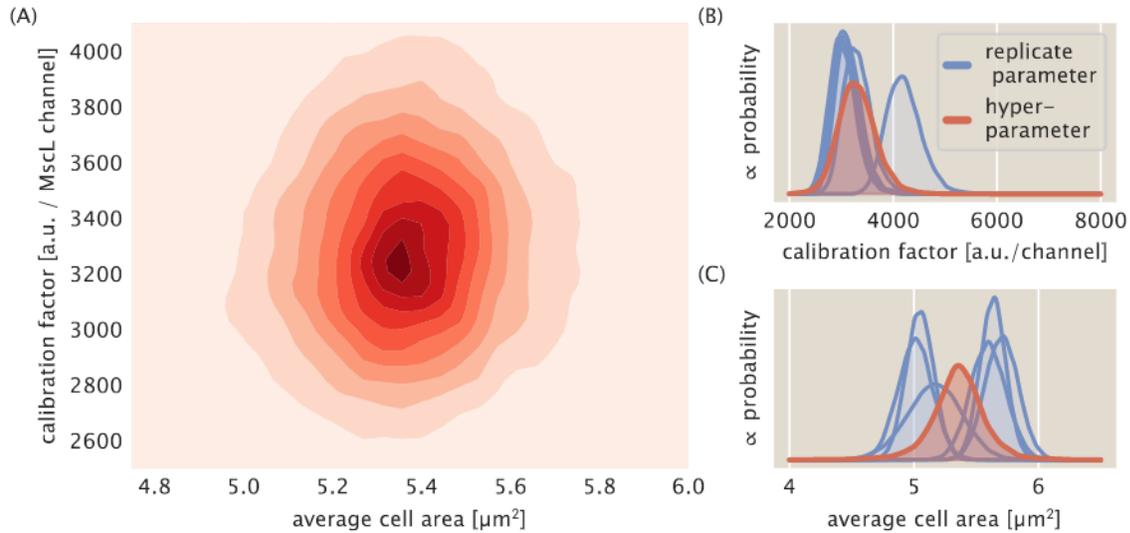

FIG 7 Posterior distributions for hyper-parameters and replicate parameters. (A) The posterior probability distribution for $\tilde{\alpha}$ and $\tilde{A}$. Probability increases from light to dark red. The replicate parameter (blue) and hyper-parameter (red) marginalized posterior probability distributions for $\alpha$ (B) and $A$ (C).

## Shock Classification

Its been previously shown that the rate of hypo-osmotic shock dictates the survival probability (4). To investigate how a single channel contributes to survival, we queried survival at several shock rates with varying MscL copy number. In the main text of this work, we separated our experiments into arbitrary bins of "fast" ( 1.0 Hz) and "slow" (< 1.0 Hz) shock rates. In this section, we discuss our rationale for coarse graining our data into these two groupings.

As is discussed in the main text and in the supplemental section *Logistic Regression*, we used a bin-free method of estimating the survival probability given the MscL channel copy number as a predictor variable. While this method requires no binning of the data, it requires a data set that sufficiently covers the physiological range of channel copy number to accurately allow prediction of survivability. Fig. 9 shows the results of the logistic regression treating each shock rate as an individual data set. The most striking feature of the plots shown in Fig. 9 is the inconsistent behavior of the predicted survivability from shock rate to shock rate. The appearance of bottle necks in the credible regions for some shock rates (0.2Hz, 0.5Hz, 2.00Hz, and 2.20 Hz) appear due to a high density of measurements within a narrow range of the channel copy number at the narrowest point in the bottle neck. While this results in a seemingly accurate prediction of the survival probability at that point, the lack of data in other copy number regimes severely limits our extrapolation outside of the copy number range of that data set. Other shock rates (0.018 Hz, 0.04 Hz, and 1.00 Hz) demonstrate completely pathological survival probability curves due to either complete survival or complete death of the population.

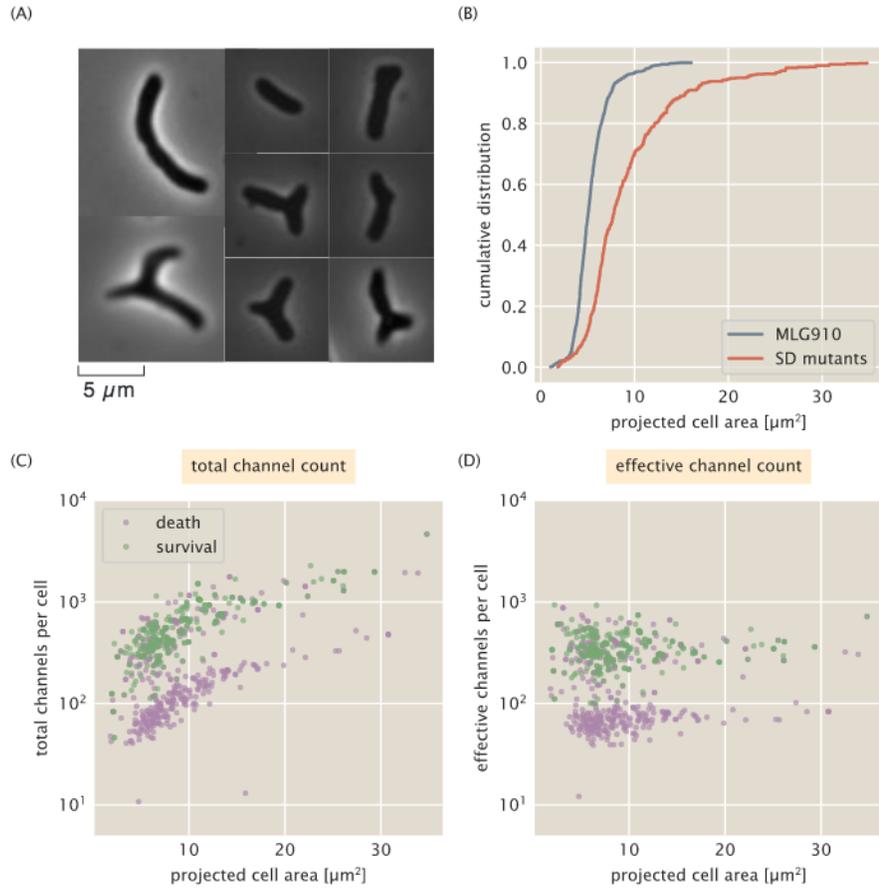

FIG 8 **Influence of area correction for Shine-Dalgarno mutants.** (A) Representative images of aberrant cell morphologies found in low-expressing Shine-Dalgarno mutants. (B) Empirical cumulative distribution of two-dimensional projected cell area for the standard candle strain MLG910 (gray line) and for all Shine-Dalgarno mutants (red line). (C) The correlation between channel copy number and cell area without the area correction. (D) The correlation between effective channel copy number and cell area with the area correction applied.

Ideally, we would like to have a wide range of MscL channel copy numbers at each shock rate shown in Fig. 9. However, the low-throughput nature of these single-cell measurements prohibits completion of this within a reasonable time frame. It is also unlikely that thoroughly dissecting the shock rate dependence will change the overall finding from our work that several hundred MscL channels are needed to convey survival under hypo-osmotic stress.

Given the data shown in Fig. 9, we can try to combine the data sets into several bins. Fig. 10 shows the data presented in Fig. 9 separated into "slow" (< 0.5 Hz, A), "intermediate" (0.5 - 1.0 Hz, B), and "fast" (> 1.0 Hz, C) shock groups. Using these groupings, the full range of MscL channel copy numbers are covered for each case, with the intermediate shock rate sparsely sampling copy numbers greater than 200 channels per cell. In all three of these cases, the same qualitative story is told – several hundred channels per cell are necessary for an appreciable level of survival when subjected to an osmotic shock. This argument is strengthened when examining the predicted survival probability by considering all shock rates as a single group, shown in Fig. 10D. This treatment tells nearly the same quantitative and qualitative story as the three rate grouping shown in this section and the two rate grouping presented in the main text. While there does appear to be a dependence on the shock rate for survival when only MscL is expressed, the effect is relatively weak with overlapping credible regions for the logistic regression across the all curves. To account for the sparse sampling of high copy numbers observed in the intermediate shock group, we split this set and partitioned the measurements into either the "slow" (< 1.0 Hz) or "fast" ( 1.0 Hz) groups presented in the main text of this work.

## Logistic Regression

In this work, we were interested in computing the survival probability under a large hypo-osmotic shock as a function of MscL channel number. As the channel copy number distributions for each Shine-Dalgarno sequence mutant were broad and overlapping, we chose to calculate the survival probability through logistic regression – a method that requires no binning of the data providing the least biased estimate of survival probability. Logistic regression is a technique that has been used in medical statistics since the late 1950's to describe diverse phenomena such as dose response curves, criminal recidivism, and survival probabilities for patients after treatment (22, 23, 50). It has also found much use in machine learning to tune a binary or categorical response given a continuous input (51–53).

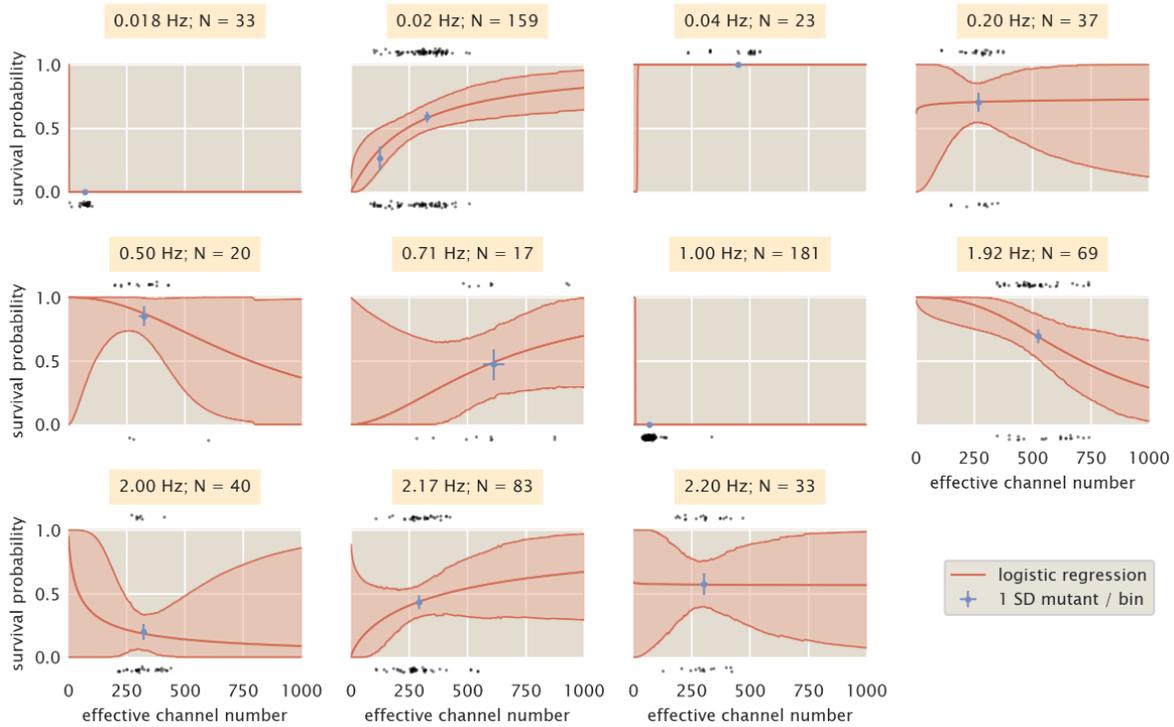

FIG 9 Binning by individual shock rates. Survival probability estimates from logistic regression (red lines) and the computed survival probability for all SD mutants subjected to that shock rate (blue points). Black points at top and bottom of each plot correspond to single cell measurements of survival (top) and death (bottom). Red shaded regions signify the 95% credible region of the logistic regression. Horizontal error bars of blue points are the standard error of the mean channel copy number. Vertical error bars of blue points correspond to the uncertainty in survival probability by observing $n$ survival events from $N$ single-cell measurements.

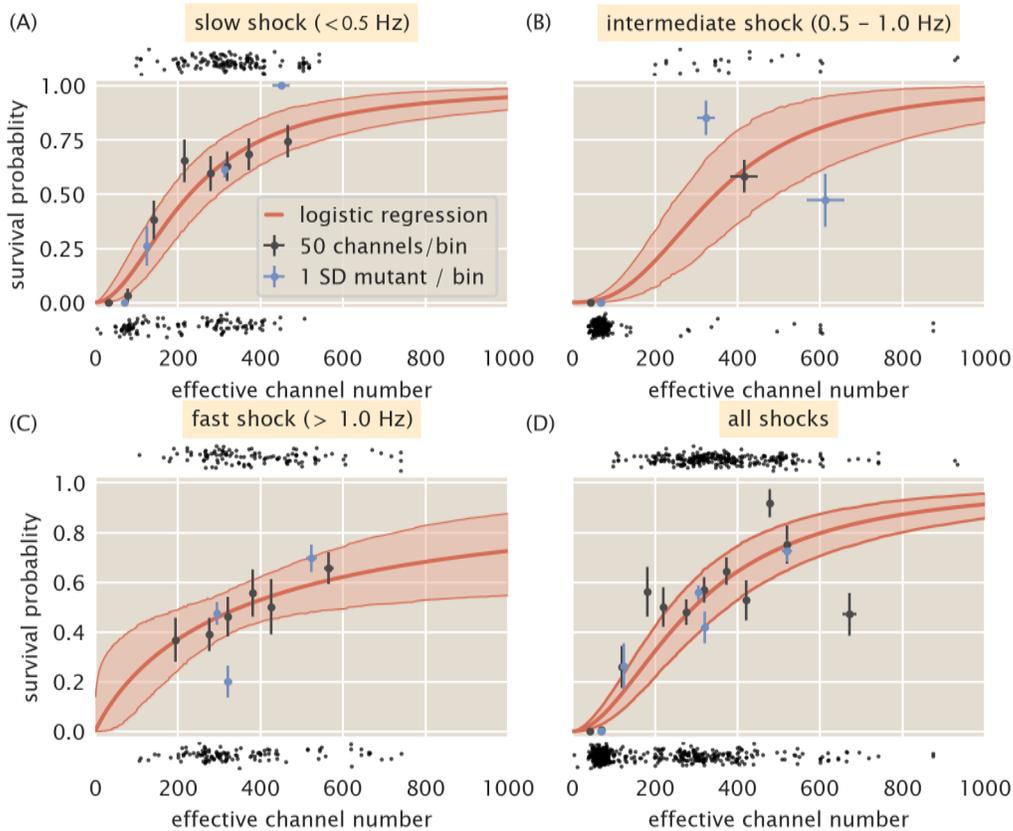

FIG 10 Coarse graining shock rates into different groups. Estimated survival probability curve for slow (A), intermediate (B), and fast (C) shock rates. (D) Estimated survival probability curve from pooling all data together, ignoring varying shock rates. Red shaded regions correspond to the 95% credible region of the survival probability estimated via logistic regression. Black points at top and bottom of each plot represent single-cell measurements of cells which survived and died, respectively. Black points and error bars represent survival probability calculations from bins of 50 channels per cell. Blue points represent the survival probability for a given Shine-Dalgarno mutant. Horizontal error bars are the standard error of the mean with at least 25 measurements and vertical error bars signifies the uncertainty in the survival probability from observing $n$ survival events out of $N$ total measurements.

In this section, we derive a statistical model for estimating the most-likely values for the coefficients $\beta_0$ and $\beta_1$ and use Bayes' theorem to provide an interpretation for the statistical meaning.

Bayesian parameter estimation of $\beta_0$ and $\beta_1$

The central challenge of this work is to estimate the probability of survival $p_s$ given only a measure of the total number of MscL channels in that cell. In other words, for a given measurement of $N_c$ channels, we want to know likelihood that a cell would survive an osmotic shock. Using Bayes' theorem, we can write a statistical model for the survival probability as

$$g(p_s \mid N_c) = \frac{f(N_c \mid p_s) g(p_s)}{f(N_c)}, \tag{31}$$

where $g$ and $f$ represent probability density functions over parameters and data, respectively. The posterior probability distribution $g(p_s \mid N_c)$ describes the probability of $p_s$ given a specific number of channels $N_c$. This distribution is dependent on the likelihood of observing $N_c$ channels assuming a value of $p_s$ multiplied by all prior knowledge we have about knowing nothing about the data, $g(s)$. The denominator $f(N_c)$ in Eq. 31 captures all knowledge we have about the available values of $N_c$, knowing nothing about the true survival probability. As this term acts as a normalization constant, we will neglect it in the following calculations for convenience.

To begin, we must come up with a statistical model that describes the experimental measurable in our experiment – survival or death. As this is a binary response, we can consider each measurement as a Bernoulli trial with a probability of success matching our probability of survival $p_s$,

$$f(s \mid p_s) = p_s^s (1 - p_s)^{1-s}, \tag{32}$$

where $s$ is the binary response of 1 or 0 for survival and death, respectively. As is stated in the introduction to this section, we decided to use a logistic function to describe the survival probability. We assume that the log-odds of survival is linear with respect to the effective channel copy number $N_c$ as

$$\log \frac{p_s}{1 - p_s} = \beta_0 + \beta_1 N_c, \tag{33}$$

where $\beta_0$ and $\beta_1$ are coefficients which describe the survival probability in the absence of channels and the increase in log-odds of survival conveyed by a single channel. The rationale behind this interpretation is presented in the following section, *A Bayesian interpretation of* $\beta_0$ *and* $\beta_1$. Using this assumption, we can solve for the survival probability $p_s$ as,

$$p_s = \frac{1}{1 + e^{-\beta_0 - \beta_1 N_c}}. \tag{34}$$

With a functional form for the survival probability, the likelihood stated in Eq. 31 can be restated as

$$f(N_c, s \mid \beta_0, \beta_1) = \left[\frac{1}{1 + e^{-\beta_0 - \beta_1 N_c}}\right]^s \left[1 - \frac{1}{1 + e^{-\beta_0 - \beta_1 N_c}}\right]^{1-s}. \tag{35}$$

As we have now introduced two parameters, $\beta_0$, and $\beta_1$, we must provide some description of our prior knowledge regarding their values. As is typically the case, we know nothing about the values for $\beta_0$ and $\beta_1$. These parameters are allowed to take any value, so long as it is a real number. Since all values are allowable, we can assume a flat distribution where any value has an equally likely probability. This value of this constant probability is not necessary for our calculation and is ignored. For a set of $k$ single-cell measurements, we can write the posterior probability distribution stated in Eq. 31 as

$$g(\beta_0, \beta_1 \mid N_c, s) = \prod_{i=1}^{n} \left[\frac{1}{1 + e^{-\beta_0 - \beta_1 N_c^{(i)}}}\right]^{s^{(i)}} \left[1 - \frac{1}{1 + e^{-\beta_0 - \beta_1 N_c^{(i)}}}\right]^{1-s^{(i)}} \tag{36}$$

Implicitly stated in Eq. 36 is absolute knowledge of the channel copy number $N_c$. However, as is described in *Standard Candle Calibration*, we must convert from a measured areal sfGFP intensity $I_A$ to a effective channel copy number,

$$N_c = \frac{I_A \tilde{A}}{\tilde{\alpha}}, \tag{37}$$

where $\tilde{A}$ is the average cell area of the standard candle strain and $\tilde{\alpha}$ is the most-likely value for the calibration factor between arbitrary units and protein copy number. In *Standard Candle Calibration*, we detailed a process for generating an estimate for the most-likely value of $\tilde{A}$ and $\tilde{\alpha}$. Given these estimates, we can include an informative prior for each value. From the Markov chain Monte Carlo samples shown in Fig. 7, the posterior distribution for each parameter is approximately Gaussian. By approximating them as Gaussian distributions, we can assign an informative prior for each as

$$g(\alpha \mid \tilde{\alpha}, \tilde{\sigma}_\alpha) \propto \frac{1}{\tilde{\sigma}_\alpha^k} \prod_{i=1}^{k} \exp\left[-\frac{(\alpha_i - \tilde{\alpha})^2}{2\tilde{\sigma}_\alpha^2}\right] \tag{38}$$

for the calibration factor for each cell and

$$g(A \mid \tilde{A}, \tilde{\sigma}_A) = \frac{1}{\tilde{\sigma}_A^k} \prod_{i=1}^{k} \exp\left[-\frac{(A_i - \tilde{A})^2}{2\tilde{\sigma}_A^2}\right], \tag{39}$$

where $\tilde{\sigma}_\alpha$ and $\tilde{\sigma}_A$ represent the variance from approximating each posterior as a Gaussian. The proportionality for each prior arises from the neglecting of normalization constants for notational convenience.

Given Eq. 35 through Eq. 39, the complete posterior distribution for estimating the most likely values of $\beta_0$ and $\beta_1$ can be written as

$$g(\beta_0, \beta_1 \mid [I_A, s], \tilde{A}, \tilde{\sigma}_A, \tilde{\alpha}, \tilde{\sigma}_\alpha) \propto \frac{1}{(\tilde{\sigma}_\alpha \tilde{\sigma}_A)^k} \prod_{i=1}^{k} \left[1 + \exp\left[-\beta_0 - \beta_1 \frac{I_{Ai} A_i}{\alpha_i}\right]\right]^{-s_i} \times$$

$$\left[1 - \left[1 + \exp\left[-\beta_0 - \beta_1 \frac{I_{Ai} A_i}{\alpha_i}\right]\right]^{-1}\right]^{1-s_i} \exp\left[-\frac{(A_i - \tilde{A})^2}{2\tilde{\sigma}_A} - \frac{(\alpha_i - \tilde{\alpha})^2}{2\tilde{\sigma}_\alpha^2}\right]. \tag{40}$$

As this posterior distribution is not solvable analytically, we used Markov chain Monte Carlo to draw samples out of this distribution, using the log of the effective channel number as described in the main text. The posterior distributions for $\beta_0$ and $\beta_1$ for both slow and fast shock rate data can be seen in Fig. 11

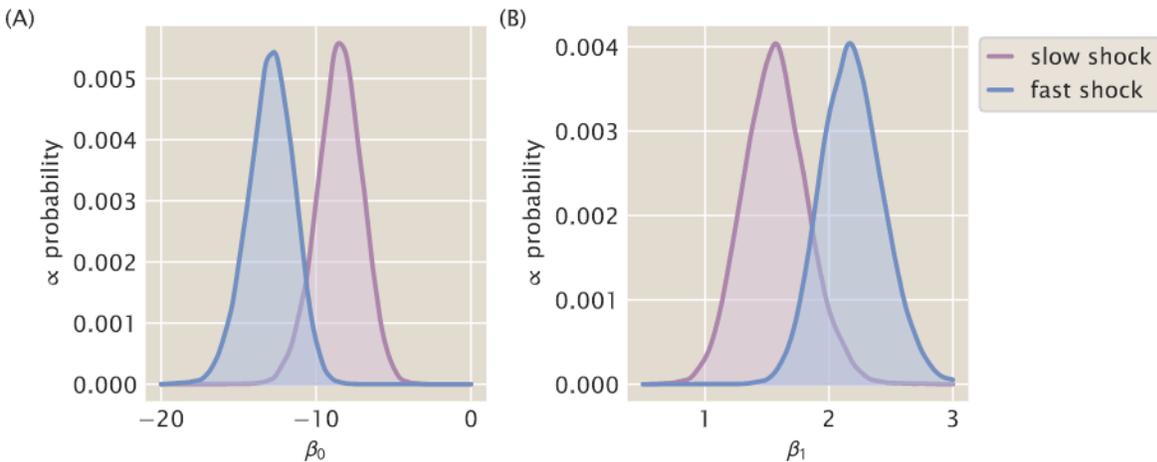

FIG 11 Posterior distributions for logistic regression coefficients evaluated for fast and slow shock rates. (A) Kernel density estimates of the posterior distribution for $\beta_0$ for fast (blue) and slow (purple) shock rates. (B) Kernel density estimates of posterior distribution for $\beta_1$.

A Bayesian interpretation of $\beta_0$ and $\beta_1$

The assumption of a linear relationship between the log-odds of survival and the predictor variable $N_c$ appears to be arbitrary and is presented without justification. However, this relationship is directly connected to the manner in which Bayes' theorem updates the posterior probability distribution upon the observation of new data. In following section, we will demonstrate this connection using the relationship between survival and channel copy number. However, this description is general and can be applied to any logistic regression model so long as the response variable is binary. This connection was shown briefly by Allen Downey in 2014 and has been expanded upon in this work (54).

The probability of observing a survival event $s$ given a measurement of $N_c$ channels can be stated using Bayes' theorem as

$$g(s \mid N_c) = \frac{f(N_c \mid s) g(s)}{f(N_c)}. \tag{41}$$

where $g$ and $f$ represent probability density functions over parameters and data respectively. The posterior distribution $g(s \mid N_c)$ is the quantity of interest and implicitly related to the probability of survival. The likelihood $g(N_c \mid s)$ tells us the probability of observing $N_c$ channels in this cell given that it survives. The quantity $g(s)$ captures all *a priori* knowledge we have regarding the probability of this cell surviving and the denominator $f(N_c)$ tells us the converse – the probability of observing $N_c$ cells irrespective of the survival outcome.

Proper calculation of Eq. 41 requires that we have knowledge of $f(N_c)$, which is difficult to estimate. While we are able to give appropriate bounds on this term, such as a requirement of positivity and some knowledge of the maximum membrane packing density, it is not so obvious to determine the distribution between these bounds. Given this difficulty, it's easier to compute the odds of survival $\mathcal{O}(s \mid N_c)$, the probability of survival $s$ relative to death $d$,

$$\mathcal{O}(s \mid N_c) = \frac{g(s \mid N_c)}{g(d \mid N_c)} = \frac{f(N_c \mid s)g(s)}{f(N_c \mid d)g(d)}, \tag{42}$$

where $f(N_c)$ is cancelled. The only stipulation on the possible value of the odds is that it must be a positive value. As we would like to equally weigh odds less than one as those of several hundred or thousand, it is more convenient to compute the log-odds, given as

$$\log \mathcal{O}(s \mid N_c) = \log \frac{g(s)}{g(d)} + \log \frac{f(N_c \mid s)}{f(N_c \mid d)}. \tag{43}$$

Computing the log-transform reveals two interesting quantities. The first term is the ratio of the priors and tells us the *a priori* knowledge of the odds of survival irrespective of the number of channels. As we have no reason to think that survival is more likely than death, this ratio goes to unity. The second term is the log likelihood ratio and tells us how likely we are to observe a given channel copy number $N_c$ given the cell survives relative to when it dies.

For each channel copy number, we can evaluate Eq. 43 to measure the log-odds of survival. If we start with zero channels per cell, we can write the log-odds of survival as

$$\log \mathcal{O}(s \mid N_c = 0) = \log \frac{g(s)}{g(d)} + \log \frac{f(N_c = 0 \mid s)}{f(N_c = 0 \mid d)}. \tag{44}$$

For a channel copy number of one, the odds of survival is

$$\log \mathcal{O}(s \mid N_c = 1) = \log \frac{g(s)}{g(d)} + \log \frac{f(N_c = 1 \mid s)}{f(N_c = 1 \mid d)}. \tag{45}$$

In both Eq. 44 and Eq. 45, the log of our *a priori* knowledge of survival versus death remains. The only factor that is changing is log likelihood ratio. We can be more general in our language and say that the log-odds of survival is increased by the difference in the log-odds conveyed by addition of a single channel. We can rewrite the log likelihood ratio in a more general form as

$$\log \frac{f(N_c \mid s)}{f(N_c \mid d)} = \log \frac{f(N_c = 0 \mid s)}{f(N_c = 0 \mid d)} + N_c \left[ \log \frac{f(N_c = 1 \mid s)}{f(N_c = 1 \mid d)} - \log \frac{f(N_c = 0 \mid s)}{f(N_c = 0 \mid d)} \right], \tag{46}$$

where we are now only considering the case in which $N_c \in [0, 1]$. The bracketed term in Eq. 46 is the log of the odds of survival given a single channel relative to the odds of survival given no channels.

Mathematically, this odds-ratio can be expressed as

$$\log \mathcal{OR}_{N_c}(s) = \log \frac{\frac{f(N_c=1\,|\,s)g(s)}{f(N_c=1\,|\,d)g(d)}}{\frac{f(N_c=0\,|\,s)g(s)}{f(N_c=0\,|\,d)g(d)}} = \log \frac{f(N_c = 1\,|\,s)}{f(N_c = 1\,|\,d)} - \log \frac{f(N_c = 0\,|\,s)}{f(N_c = 0\,|\,d)}. \tag{47}$$

Eq. 47 is mathematically equivalent to the bracketed term shown in Eq. 46.

We can now begin to staple these pieces together to arrive at an expression for the log odds of survival. Combining Eq. 46 with Eq. 43 yields

$$\log \mathcal{O}(s\,|\,N_c) = \log \frac{g(s)}{g(d)} + \log \frac{f(N_c = 0\,|\,s)}{f(N_c = 0\,|\,d)} + N_c \left[ \log \frac{f(N_c = 1\,|\,s)}{f(N_c = 1\,|\,d)} - \log \frac{f(N_c = 0\,|\,s)}{f(N_c = 0\,|\,d)} \right]. \tag{48}$$

Using our knowledge that the bracketed term is the log odds-ratio and the first two times represents the log-odds of survival with $N_c = 0$, we conclude with

$$\log \mathcal{O}(s\,|\,N_c) = \log \mathcal{O}(s\,|\,N_c = 0) + N_c \log \mathcal{OR}_{N_c}(s). \tag{49}$$

This result can be directly compared to Eq. 1 presented in the main text,

$$\log \frac{p_s}{1 - p_s} = \beta_0 + \beta_1 N_c, \tag{50}$$

which allows for an interpretation of the seemingly arbitrary coefficients $\beta_0$ and $\beta_1$. The intercept term, $\beta_0$, captures the log-odds of survival with no MscL channels. The slope, $\beta_1$, describes the log odds-ratio of survival which a single channel relative to the odds of survival with no channels at all. While we have examined this considering only two possible channel copy numbers (1 and 0), the relationship between them is linear. We can therefore generalize this for any MscL copy number as the increase in the log-odds of survival is constant for addition of a single channel.

Other properties as predictor variables

The previous two sections discuss in detail the logic and practice behind the application of logistic regression to cell survival data using only the effective channel copy number as the predictor of survival. However, there are a variety of properties that could rightly be used as predictor variables, such as cell area and shock rate. As is stipulated in our standard candle calibration, there should be no correlation between survival and cell area. Fig. 12A and B show the logistic regression performed on the cell area. We see for both slow and fast shock groups,there is little change in survival probability with changing cell area and the wide credible regions allow for both positive and negative correlation between survival and area. The appearance of a bottle neck in the notably wide credible regions is a result of a large fraction of the measurements being tightly distributed about a mean value. Fig. 12C shows the predicted survival probability as a function of the the shock rate. There is a slight decrease in survivability as a function of increasing shock rate, however the width of the credible region allows for slightly positive or slightly negative correlation. While we have presented logistic regression in this section as a one-dimensional method, Eq. 33 can be generalized to $n$ predictor variables $x$ as

$$\log \frac{p_s}{1 - p_s} = \beta_0 + \sum_i^n \beta_i x_i. \tag{51}$$

Using this generalization, we can use both shock rate and the effective channel copy number as predictor variables. The resulting two-dimensional surface of survival probability is shown in Fig. 12D. As is suggested by Fig. 12C, the magnitude of change in survivability as the shock rate is increased is smaller than that along increasing channel copy number, supporting our conclusion that for MscL alone, the copy number is the most important variable in determining survival.

TABLE 2 *Escherichia coli* strains used in this work.

| Strain name | Genotype | Reference |
| --- | --- | --- |
| MJF641 | Frag1, Δ*mscL::cm*, Δ*mscS*, Δ*mscK::kan*, Δ*ybdG::apr*, Δ*ynaI*, Δ*yjeP*, Δ*ybiO*, *ycjM::Tn10* | (5) |
| MLG910 | MG1655, Δ*mscL* ::φ*mscL-sfGFP*, Δ*galK::kan*, Δ*lacI*, Δ*lacZY A* | (3) |
| D6LG-Tn10 | Frag1, Δ*mscL* ::φ*mscL-sfGFP*, Δ*mscS*, Δ*mscK::kan*, Δ*ybdG::apr*, Δ*ynaI*, Δ*yjeP*, Δ*ybiO*, *ycjM::Tn10* | This work |
| D6LG (SD0) | Frag1, Δ*mscL* ::φ*mscL-sfGFP*, Δ*mscS*, Δ*mscK::kan*, Δ*ybdG::apr*, Δ*ynaI*, Δ*yjeP*, Δ*ybiO* | This work |
| XTL298 | CC4231, *araD:: tetA-sacB-amp* | (43) |
| D6LTetSac | Frag1, *mscL-sfGFP:: tetA-sacB*, Δ*mscS*, Δ*mscK::kan*, Δ*ybdG::apr*, Δ*ynaI*, Δ*yjeP*, Δ*ybiO* | This work |
| D6LG (SD1) | Frag1, Δ*mscL* ::φ*mscL-sfGFP*, Δ*mscS*, Δ*mscK::kan*, Δ*ybdG::apr*, Δ*ynaI*, Δ*yjeP*, Δ*ybiO* | This work |
| D6LG (SD2) | Frag1, Δ*mscL* ::φ*mscL-sfGFP*, Δ*mscS*, Δ*mscK::kan*, Δ*ybdG::apr*, Δ*ynaI*, Δ*yjeP*, Δ*ybiO* | This work |
| D6LG (SD4) | Frag1, Δ*mscL* ::φ*mscL-sfGFP*, Δ*mscS*, Δ*mscK::kan*, Δ*ybdG::apr*, Δ*ynaI*, Δ*yjeP*, Δ*ybiO* | This work |
| D6LG (SD6) | Frag1, Δ*mscL* ::φ*mscL-sfGFP*, Δ*mscS*, Δ*mscK::kan*, Δ*ybdG::apr*, Δ*ynaI*, Δ*yjeP*, Δ*ybiO* | This work |
| D6LG (12SD2) | Frag1, Δ*mscL* ::φ*mscL-sfGFP*, Δ*mscS*, Δ*mscK::kan*, Δ*ybdG::apr*, Δ*ynaI*, Δ*yjeP*, Δ*ybiO* | This work |
| D6LG (16SD0) | Frag1, Δ*mscL* ::φ*mscL-sfGFP*, Δ*mscS*, Δ*mscK::kan*, Δ*ybdG::apr*, Δ*ynaI*, Δ*yjeP*, Δ*ybiO* | This work |

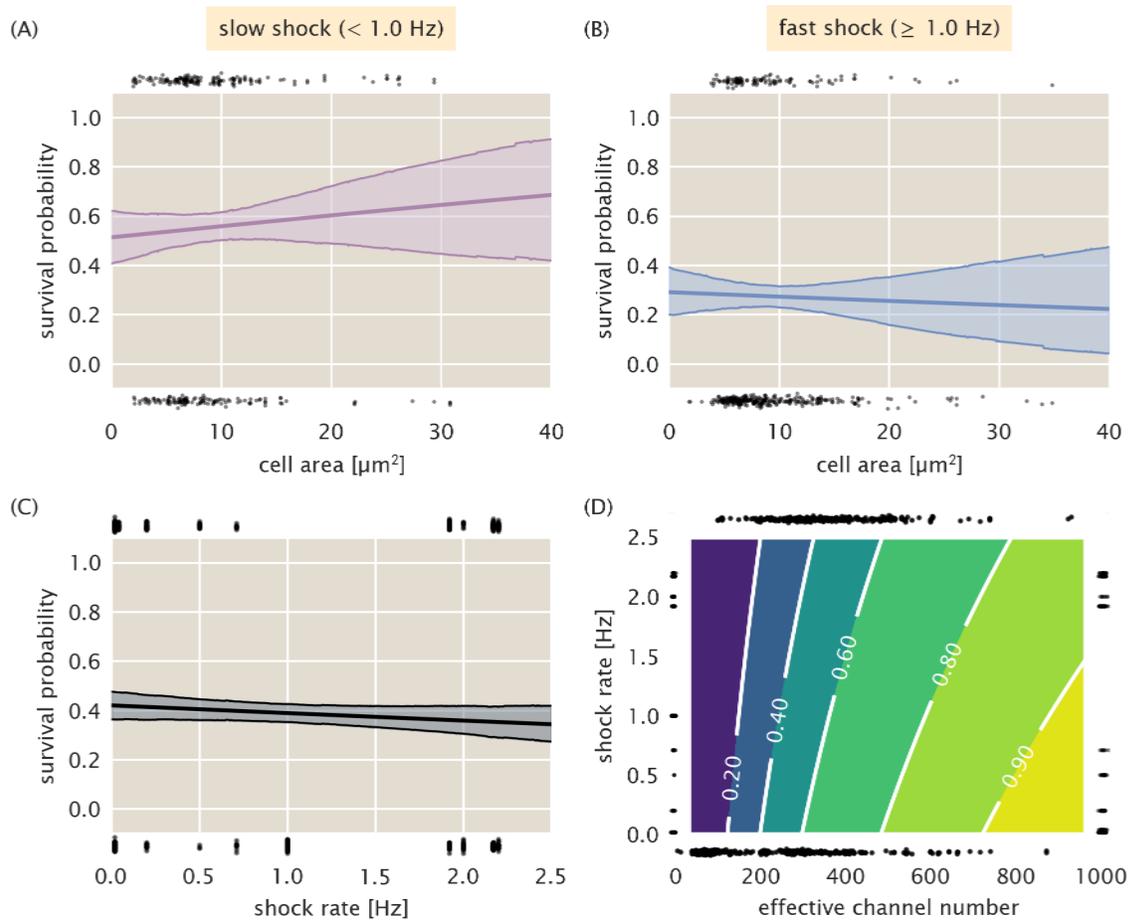

FIG 12 Survival probability estimation using alternative predictor variables. (A) Estimated survival probability as a function of cell area for the slow shock group. (B) Estimated survival probability as a function of cell area for the fast shock group. (C) Estimated survival probability as a function shock rate. Black points at top and bottom of plots represent single-cell measurements of cells who survived and perished, respectively. Shaded regions in (A) - (C) represent the 95% credible region. (D) Surface of estimated survival probability using both shock rate and effective channel number as predictor variables. Black points at left and right of plot represent single-cell measurements of cells which survived and died, respectively, sorted by shock rate. Points at top and bottom of plot represent survival and death sorted by their effective channel copy number. Labeled contours correspond to the survival probability.

TABLE 3 Oligonucleotide sequences used in this work. Bold and italics correspond to Shine-Dalgarno sequence modifications and `AT` hairpin insertion modifications, respectively. Double bar `||` indicates a transposon insertion site.

| Primer Name | Sequence (5' → 3') |
| --- | --- |
| *Tn10delR* | taaagccaacggcatccaggcggacatactcagca\|\|cctttcgcaaggtaacagagtaaaacatccaccat |
| *MscLSPSac* | gaaaatggcttaacatttgttagacttatggttgtcggcttcat**agggag**TCCTAATTTTTGTTGACACTCTATC |
| *MscLSPSacR* | accacgttcccgcgcatcgcaaattcgcgaaattctttaataatgctcatATCAAAGGGAAAACTGTCCATA |
| *MscL-SD1R* | atcgcaaattcgcgaaattctttaataatgctcatgttatt**ctcctc**atgaagccgacaaccataagtctaacaaa |
| *MscL-SD2R* | atcgcaaattcgcgaaattctttaataatgctcat**gttatt tcccct**atgaagccgacaaccataagtctaacaaa |
| *MscL-SD4R* | atcgcaaattcgcgaaattctttaataatgctcat**gttatt cctgct**atgaagccgacaaccataagtctaacaaa |
| *MscL-SD6R* | atcgcaaattcgcgaaattctttaataatgctcat**gttatt gctcgt**atgaagccgacaaccataagtctaacaaa |
| *MscL-12SD2R* | atcgcaaattcgcgaaattctttaataatgctcat*atatatatatat* **tcccct**atgaagccgacaaccataagtctaacaaa |
| *MscL-16SD0R* | atcgcaaattcgcgaaattctttaataatgctcat*atatatatatatatat* **ctccct**atgaagccgacaaccataagtctaacaaa |

## References

1. Martinac B, Buechner M, Delcour AH, Adler J, Kung C. 1987. Pressure-sensitive ion channel in *Escherichia coli*. Proc Natl Acad Sci U S A 84:2297–301.

2. Bavi N, Cortes DM, Cox CD, Rohde PR, Liu W, Deitmer JW, Bavi O, Strop P, Hill AP, Rees D, Corry B, Perozo E, Martinac B. 2016. The role of MscL amphipathic N terminus indicates a blueprint for bilayer-mediated gating of mechanosensitive channels. Nature Communications 7:11984.

3. Bialecka-Fornal M, Lee HJ, DeBerg HA, Gandhi CS, Phillips R. 2012. Single-Cell Census of

Mechanosensitive Channels in Living Bacteria. PLoS ONE 7:e33077.

4. Bialecka-Fornal M, Lee HJ, Phillips R. 2015. The Rate of Osmotic Downshock Determines the Survival Probability of Bacterial Mechanosensitive Channel Mutants. Journal of Bacteriology 197:231–237.

5. Edwards MD, Black S, Rasmussen T, Rasmussen A, Stokes NR, Stephen TL, Miller S, Booth IR. Jul-Aug 20122012. Characterization of three novel mechanosensitive channel activities in *Escherichia coli*. Channels (Austin) 6:272–81.

6. Naismith JH, Booth IR. 2012. Bacterial mechanosensitive channels–MscS: Evolution's solution to creating sensitivity in function. Annu Rev Biophys 41:157–77.

7. Ursell T, Phillips R, Kondev J, Reeves D, Wiggins PA. 2008. The role of lipid bilayer mechanics in mechanosensation, pp. 37–70. *In* Kamkin, A, Kiseleva, I (eds.), Mechanosensitivity in cells and tissues 1: Mechanosensitive ion channels. Springer-Verlag.

8. van den Berg J, Galbiati H, Rasmussen A, Miller S, Poolman B. 2016. On the mobility, membrane location and functionality of mechanosensitive channels in *Escherichia coli*. Scientific Reports 6.

9. Cruickshank CC, Minchin RF, Le Dain AC, Martinac B. 1997. Estimation of the pore size of the large-conductance mechanosensitive ion channel of *Escherichia coli*. Biophysical Journal 73:1925–1931.

10. Haswell ES, Phillips R, Rees DC. 2011. Mechanosensitive Channels: What Can They Do and How Do They Do It? Structure 19:1356–1369.

11. Louhivuori M, Risselada HJ, van der Giessen E, Marrink SJ. 2010. Release of content through mechano-sensitive gates in pressurized liposomes. Proc Natl Acad Sci U S A 107:19856–60.

12. Milo R, Jorgensen P, Moran U, Weber G, Springer M. 2010. BioNumbersthe database of key numbers in molecular and cell biology. Nucleic Acids Research 38:D750–D753.

13. Booth IR, Edwards MD, Murray E, Miller S. 2005. The role of bacterial ion channels in cell physiology, pp. 291–312. *In* Kubalsi, A, Martinac, B (eds.), Bacterial Ion Channels and Their Eukaryotic Homologs. American Society for Microbiology, Washington DC.

14. Hase CC, Minchin RF, Kloda A, Martinac B. 1997. Cross-linking studies and membrane localization and assembly of radiolabelled large mechanosensitive ion channel (MscL) of *Escherichia coli*. Biochem Biophys Res Commun 232:777–82.

15. Schmidt A, Kochanowski K, Vedelaar S, Ahrné E, Volkmer B, Callipo L, Knoops K, Bauer M, Aebersold R, Heinemann M. 2016. The quantitative and condition-dependent *Escherichia coli* proteome. Nature Biotechnology 34:104–110.

16. Soufi B, Krug K, Harst A, Macek B. 2015. Characterization of the E. coli proteome and its modifica-

tions during growth and ethanol stress. Frontiers in Microbiology 6.

17. Stokes NR, Murray HD, Subramaniam C, Gourse RL, Louis P, Bartlett W, Miller S, Booth IR. 2003. A role for mechanosensitive channels in survival of stationary phase: Regulation of channel expression by RpoS. Proceedings of the National Academy of Sciences 100:15959–15964.

18. Norman C, Liu ZW, Rigby P, Raso A, Petrov Y, Martinac B. 2005. Visualisation of the mechanosensitive channel of large conductance in bacteria using confocal microscopy. Eur Biophys J 34:396–402.

19. Espah Borujeni A, Channarasappa AS, Salis HM. 2014. Translation rate is controlled by coupled trade-offs between site accessibility, selective RNA unfolding and sliding at upstream standby sites. Nucleic Acids Research 42:2646–2659.

20. Salis HM, Mirsky EA, Voigt CA. 2009. Automated design of synthetic ribosome binding sites to control protein expression. Nature Biotechnology 27:946–950.

21. Elowitz MB, Levine AJ, Siggia ED, Swain PS. 2002. Stochastic gene expression in a single cell. Science 297:1183–6.

22. Anderson RP, Jin R, Grunkemeier GL. 2003. Understanding logistic regression analysis in clinical reports: An introduction. The Annals of Thoracic Surgery 75:753–757.

23. Mishra V, Skotak M, Schuetz H, Heller A, Haorah J, Chandra N. 2016. Primary blast causes mild, moderate, severe and lethal TBI with increasing blast overpressures: Experimental rat injury model. Scientific Reports 6:26992.

24. Feeling-Taylor AR, Yau S-T, Petsev DN, Nagel RL, Hirsch RE, Vekilov PG. 2004. Crystallization Mechanisms of Hemoglobin C in the R State. Biophysical Journal 87:2621–2629.

25. Finch JT, Perutz MF, Bertles JF, Dobler J. 1973. Structure of Sickled Erythrocytes and of Sickle-Cell Hemoglobin Fibers. Proceedings of the National Academy of Sciences 70:718–722.

26. Perutz MF, Mitchison JM. 1950. State of Hæmoglobin in Sickle-Cell Anæmia. Nature 166:677–679.

27. Berg H, Purcell E. 1977. Physics of chemoreception. Biophysical Journal 20:193–219.

28. Colin R, Sourjik V. 2017. Emergent properties of bacterial chemotaxis pathway. Current Opinion in Microbiology 39:24–33.

29. Krembel A, Colin R, Sourjik V. 2015. Importance of Multiple Methylation Sites in *Escherichia coli* Chemotaxis. PLoS ONE 10.

30. Krembel AK, Neumann S, Sourjik V. 2015. Universal Response-Adaptation Relation in Bacterial

Chemotaxis. Journal of Bacteriology 197:307–313.

31. Sourjik V, Berg HC. 2002. Receptor sensitivity in bacterial chemotaxis. Proceedings of the National Academy of Sciences 99:123–127.

32. Liu F, Morrison AH, Gregor T. 2013. Dynamic interpretation of maternal inputs by the Drosophila segmentation gene network. Proceedings of the National Academy of Sciences of the United States of America 110:6724–6729.

33. Lovely GA, Brewster RC, Schatz DG, Baltimore D, Phillips R. 2015. Single-molecule analysis of RAG-mediated V(D)J DNA cleavage. Proceedings of the National Academy of Sciences 112:E1715–E1723.

34. Schatz DG, Baltimore D. 2004. Uncovering the V(D)J recombinase. Cell 116:S103–S108.

35. Schatz DG, Ji Y. 2011. Recombination centres and the orchestration of V(D)J recombination. Nature Reviews Immunology 11:251–263.

36. Herbig U, Jobling WA, Chen BP, Chen DJ, Sedivy JM. 2004. Telomere Shortening Triggers Senescence of Human Cells through a Pathway Involving ATM, p53, and p21CIP1, but Not p16INK4a. Molecular Cell 14:501–513.

37. Victorelli S, Passos JF. 2017. Telomeres and Cell Senescence - Size Matters Not. EBioMedicine 21:14–20.

38. Booth IR. 2014. Bacterial mechanosensitive channels: Progress towards an understanding of their roles in cell physiology. Current Opinion in Microbiology 18:16–22.

39. Li G-W, Burkhardt D, Gross C, Weissman JS. 2014. Quantifying Absolute Protein Synthesis Rates Reveals Principles Underlying Allocation of Cellular Resources. Cell 157:624–635.

40. Blount P, Sukharev SI, Moe PC, Martinac B, Kung C. 1999. Mechanosensitive channels of bacteria. Methods in Enzymology 294:458–482.

41. Schumann U, Edwards MD, Rasmussen T, Bartlett W, van West P, Booth IR. 2010. YbdG in *Escherichia coli* is a threshold-setting mechanosensitive channel with MscM activity. Proc Natl Acad Sci U S A 107:12664–9.

42. Sharan SK, Thomason LC, Kuznetsov SG, Court DL. 2009. Recombineering: A homologous recombination-based method of genetic engineering. Nat Protoc 4:206–23.

43. Li X-t, Thomason LC, Sawitzke JA, Costantino N, Court DL. 2013. Positive and negative selection using the tetA-sacB cassette: Recombineering and P1 transduction in *Escherichia coli*. Nucleic acids research 41:e204–e204.

44. Bochner BR, Huang H-C, Schieven GL, Ames BN. 1980. Positive selection for loss of tetracycline
43

resistance. Journal of bacteriology 143:926–933.

45. Carpenter B, Gelman A, Hoffman MD, Lee D, Goodrich B, Betancourt M, Brubaker M, Guo J, Li P, Riddell A. 2017. Stan : A Probabilistic Programming Language. Journal of Statistical Software 76.

46. Chure G, Lee HJ, Phillips R. 2018. Image data for "Connecting the dots between mechanosensitive channel abundance, osmotic shock, and survival at single-cell resolution" accessible through DOI: 10.22002/D1.941.

47. Chure G, Lee HJ, Phillips R. 2018. MCMC chains generated in "Connecting the dots between mechanosensitive channel abundance, osmotic shock, and survival at single-cell resolution" accessible through DOI 10.22002/D1.942.

48. Chure G, Lee HJ, Phillips R. 2018. Github repository for "Connecting the dots between mechanosensitive channel abundance, osmotic shock, and survival at single-cell resolution" accessible through DOI: 10.5281/zenodo.1252524.

49. Sivia D, Skilling J. 2006. Data Analysis: A Bayesian TutorialSecond Edition. Oxford University Press, Oxford, New York.

50. Stahler GJ, Mennis J, Belenko S, Welsh WN, Hiller ML, Zajac G. 2013. Predicting Recidivism For Released State Prison Offenders. Criminal justice and behavior 40:690–711.

51. Cheng W, Hüllermeier E. 2009. Combining instance-based learning and logistic regression for multilabel classification. Machine Learning 76:211–225.

52. Dreiseitl S, Ohno-Machado L. 2002. Logistic regression and artificial neural network classification models: A methodology review. Journal of Biomedical Informatics 35:352–359.

53. Ng AY, Jordan MI. On Discriminative vs. Generative Classifiers: A comparison of logistic regression and naive Bayes 8.

54. Downey A. 2014. Probably Overthinking It: Bayes's theorem and logistic regression. Probably Overthinking It.